\documentclass[aps,prl,10pt,twocolumn,amsmath,amssymb,floatfix,superscriptaddress,longbibliography]{revtex4-2}

\usepackage{graphicx} 
\usepackage{bm} 

\usepackage{hyperref}

\usepackage{braket,bbold,color,enumerate}
\usepackage{float}

\usepackage{geometry}
\usepackage{amsmath,amssymb,xcolor}
\usepackage{amsfonts}
\usepackage{amsmath}
\usepackage{mathtools}
\usepackage{dsfont}
\usepackage[capitalize]{cleveref}

\newcommand{\cdag}{c^{\dag}}

\newcommand{\im}{\mathrm{i}}
\newcommand{\id}{\mathbb{1}}

\newcommand{\Chat}{\hat{\mathcal{C}}}

\newcommand{\Cabk}{\mathcal{C}_{ab}(k)}

\newcommand{\SWAP}{\hat{\mathcal{S}}}

\usepackage{xcolor}

\begin{document}

\title{Topology from Decoherence}

\author{Alexandre Chaduteau}
\affiliation{Blackett Laboratory, Imperial College London, London SW7 2AZ, United Kingdom}

\author{Derek K. K. Lee}
\affiliation{Blackett Laboratory, Imperial College London, London SW7 2AZ, United Kingdom}

\author{Frank Schindler}
\affiliation{Blackett Laboratory, Imperial College London, London SW7 2AZ, United Kingdom}

\author{Abhinav Prem}
\affiliation{Physics Program, Bard College, 30 Campus Road, Annandale-on-Hudson, New York 12504, USA}

\begin{abstract}
Decoherence is conventionally regarded as an obstacle to realizing topological quantum phases. This has motivated extensive efforts to suppress noise in candidate topological materials and devices. Here, we show that decoherence can instead induce topological phenomena. We demonstrate this in a lattice system subject to environment-induced dephasing. The noise-averaged dynamics, governed by an interacting quantum master equation, realize a topological phase characterized by a winding number and the non-Hermitian skin effect. The dynamical consequence is striking: the correlated nature of the stochastic noise yields asymmetric diffusion, whose direction is fixed by the winding number and is reversible only through a topological phase transition. This effect is induced purely by interactions, distinguishing it from previous studies of free, effectively single-particle systems. It also disappears upon postselecting measurement outcomes, confirming that it is a genuinely open-system phenomenon with no effective Hamiltonian description. Remarkably, the model remains analytically tractable. Our results establish correlated quantum noise as a route to topology in open many-body systems, beyond free-particle and non-Hermitian Hamiltonian paradigms.
\end{abstract}
\maketitle


\emph{Introduction---}
Topological phases of quantum matter owe their appeal to their remarkable rigidity~\cite{RevModPhys.89.041004,Gilbert2021TopologicalElectronics}: quantized responses and protected boundary modes that persist under arbitrary local perturbations. This robustness has motivated a broad effort to harness topological protection in quantum technologies, from metrology~\cite{PhysRevLett.45.494} to quantum computation~\cite{RevModPhys.80.1083}. However, these systems are treated as isolated. Realistic quantum systems inevitably interact with their environment, leading to dissipation and decoherence. Whether topological phenomena can survive or even emerge under such open-system conditions is therefore a question of both fundamental and practical importance.

A natural, minimal framework for describing open quantum dynamics is
the Lindblad master equation~\cite{lindblad1976generators, gorini1976completely}, which encompasses paradigmatic models such as quantum Brownian motion~\cite{PhysRevA.94.042123}. In this setting, mixed-state and Lindbladian topology has recently come into focus~\cite{lieu2020tenfold, yang2022lse, porras2022bridge, Chaduteau_2025, zhou2022non, degroot2022,ma2023aspt,sohal2025imto,ellison2025imto, 8vtp-khnq, PhysRevX.15.021016}. This program becomes timely for applications, given the exquisite control over driving and dissipation now achievable in quantum materials and simulators~\cite{RevModPhys.86.153}. However, analytic understanding of topology in Lindbladians remains almost entirely confined to non-interacting systems, admitting an effective single-particle description. It remains an outstanding question whether noisy dynamics can not only protect, but instead generate, topological phenomena in interacting open systems.
\begin{figure*}[tp]
    \centering
    \includegraphics[width=1.2\columnwidth]{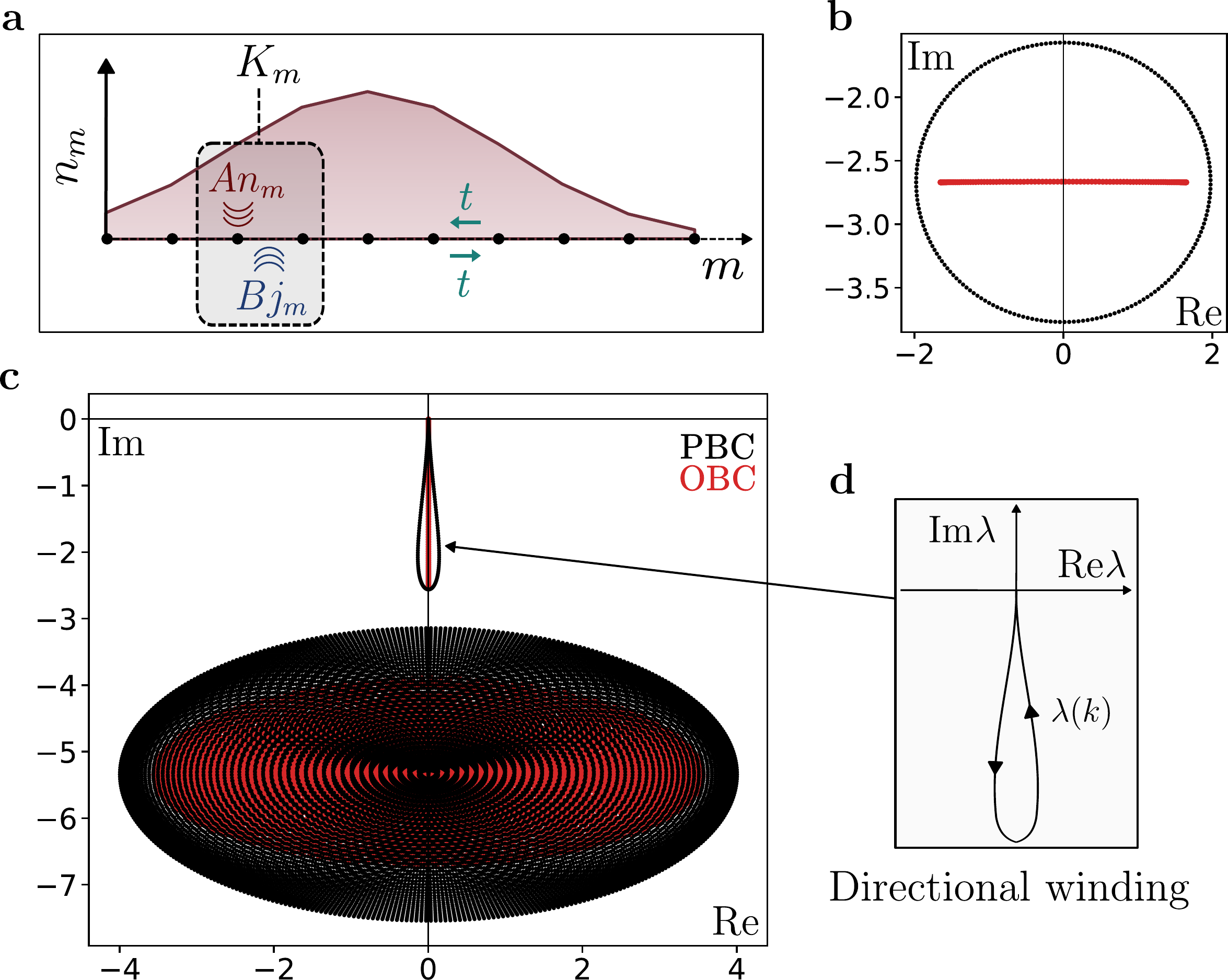}
    \caption{\small{\textbf{Decoherence-induced topology in a lattice model.} 
    \textbf{a.} Schematic of the proposed model [see Eq.~\eqref{eq:interacting_model}]. The Hamiltonian drives reciprocal coherent hopping with strength $t$. The jumps are $K_m = An_m + Bj_m$, with on-site density $n_m$ at site $m$ and current $j_{m}$ between sites $m$ and $m+1$.
    \textbf{b.} Complex eigenspectrum of the single-particle sector generator $\Theta$ for $t=1, A=2.2, B=0.5$ and $N=100$. In PBC (black), the spectrum displays a point gap~\cite{kawabata_38_fold_way, PhysRevLett.124.086801}, causing a non-Hermitian skin effect in the single-particle eigenstates of the Lindbladian.
    \textbf{c.} Eigenspectrum of the two-particle sector generator $\Chat$ for the same parameters. We observe two parts: the ``continuous" spectrum~\cite{Hislop1996} corresponding to postselected dynamics, and the ``discrete" part caused by the quantum jumps. The former is understood from $\Theta$, while the latter exhibits a ``tear"-shaped point gap in PBC (black).
    \textbf{d.} Magnification of the tear-shaped point gap. The complex spectrum winds around a base point (arrow indicates orientation) against momentum $k$ in the Brillouin zone. This is the topological feature underlying the asymmetric diffusion of Fig.~\ref{fig:dynamics_figure}.
    }}
    \label{fig:model_spectrum}
\end{figure*}

We answer this question affirmatively, presenting a model of \textit{decoherence-induced topology}: a one-dimensional lattice with stochastic quantum jumps~\cite{Jacobs01092006, Murugan2017, PhysRevX.11.031015} involving correlated density and current measurements. The noise-averaged dynamics are governed by a Lindbladian whose spectrum exhibits non-trivial topology. Unlike prior studies of topology in Lindbladians, the effect is induced by interactions with the environment but remains analytically tractable. We develop a method resolving this decoherence-induced topology with polynomial complexity in system size. This is markedly distinct from interacting Hamiltonians, where analytical control is elusive.

In the context of dissipation, non-Hermitian Hamiltonians and their topological properties have been extensively studied~\cite{kawabata_38_fold_way, PhysRevLett.124.086801, PhysRevB.105.165137, PhysRevX.8.031079}, but describe only postselected dynamics~\cite{Chaduteau_2025}. 
The topology in our model instead resides in the full Lindbladian, free of postselection and describing the entire many-body system. This obviates the challenges associated with realizing postselected or monitored dynamics and points toward accessible experimental realizations~\cite{Bloch2012QuantumSimulations, PhysRevA.82.063605, Dalibard_1985,PhysRevLett.127.070402,PhysRevLett.123.080601}.

A crisp dynamical consequence of the many-body topology is asymmetric diffusion in the approach to the steady-state, with directionality dictated by the topological invariant. In typical models of decoherence, stochastic noise produces symmetric diffusion with no directional bias~\cite{PhysRevLett.127.070402,Eisler_2011,PhysRevLett.123.080601,PhysRevA.87.012108}. Here, the directionality emerges from the correlations between the quantum jumps. Our results show that correlated quantum noise can itself serve as a topological resource, generating dynamical topological phenomena without counterpart in coherent dynamics. This sharply contrasts with previous work studying robustness of conventional topological phases under decoherence (see Refs.~\onlinecite{Ho_2014, Lee_SPT_decoherence_2025, 8vtp-khnq} and references therein), as well as work on dissipatively engineered steady states~\cite{bardyn2013top}.

\emph{The model---}
Consider a one-dimensional (1D) system of (spinless) fermions with annihilation/creation operators $c_i/\cdag_i,\; i=1\ldots N$ for some integer $N$. These obey canonical anti-commutation relations $\{c_i,c_j\} = 0, \:\{\cdag_i,c_j\}=\delta_{ij}$, where curly brackets denote the anti-commutator and $\delta_{ij}$ is the Kronecker delta symbol. We restrict ourselves to quadratic, particle-number conserving (Hermitian) Hamiltonians $H = \sum_{i,j=1}^N\mathcal{H}_{ij}\cdag_ic_j$, where $\mathcal{H}$ is a Hermitian $N\times N$ matrix. Similarly, we define a family of quadratic Hermitian operators $K_m$, indexed by $m$, of the form $K_m = \sum_{ij}(\mathcal{K}_m)_{ij}\cdag_ic_j$.
Of particular interest for dephasing models~\cite{Znidaric_2010_1, Znidaric_2010_2, PhysRevLett.117.137202, 8vtp-khnq} are densities $n_m=\cdag_m c_m$ and currents $j_{m}=\im (\cdag_mc_{m+1}-\cdag_{m+1}c_m)$, in which case the index $m=1 \dots N$ corresponds to the fermionic modes. Our model with quantum jumps is depicted in Fig.~\ref{fig:model_spectrum}(a), with
\begin{equation}\label{eq:interacting_model}
    \begin{aligned}
        H &= t\sum_{i=1}^N (\cdag_{i+1}c_i + \cdag_{i}c_{i+1}), \\
        K_m &= A n_m + Bj_{m},
    \end{aligned}
\end{equation}
where $i$ and $m$ are now spatial indices, and $t, A,B$ are real parameters.
We consider both open/periodic boundary conditions (OBC/PBC); the PBC version identifies $c_{N+1}\equiv c_1$. The $K_m$ satisfy the locality condition $(\mathcal{K}_m)_{ij}=0$ for all $i,j$ such that $|i-m|>d$ or $|j-m|>d$ where $d=1$ here. They are translationally invariant, in that $(\mathcal{K}_m)_{ij}=(\mathcal{K}_{m+1})_{i+1,j+1}$.

We consider time ($\tau$) evolution of a density matrix $\rho(\tau)$ via the Lindblad equation~\cite{gorini1976completely, lindblad1976generators, PhysRevA.87.012108, Barthel_2022, Zunkovic_2014, Eisler_2011}
\begin{equation}\label{eq:lindbladian_quadratic_hermitian_jumps}
    \im\frac{\mathrm{d}\rho}{\mathrm{d}\tau} = \mathcal{L}(\rho)=\left[H,\rho\right] - \frac{\im}{2} \sum_m \left[K_m,\left[K_m, \rho\right]\right],
\end{equation}
where $\mathcal{L}$ is the Lindbladian. Physically, this master equation describes noise-averaged dynamics generated by $H$ alongside stochastic driving by the jumps $K_m$, under standard white-noise assumptions~\cite{Jacobs01092006}. The Lindbladian superoperator contains terms quartic in the fermions and hence describes an interacting open system. For Hermitian jumps, the steady state $\rho_S$ (satisfying $\mathrm{d}\rho_S/\mathrm{d}\tau = 0$) is necessarily proportional to the identity~\cite{PhysRevA.87.012108}, i.e. the infinite-temperature mixed state. Two-particle observables, such as densities $n_i$ and currents $j_i$, are obtained from the correlation matrix $C_{ij} := \mathrm{Tr}(\rho \cdag_i c_j)$. 
In the Supplementary Material (SM)~\footnote{See the Supplementary Material for detailed derivations, including equations of motion, momentum-space expressions, and a detailed study of the skin effect.}, we obtain its equation of motion as
\begin{equation}\label{eq:CM_EOM}
\begin{aligned}
    \im\frac{\mathrm{d}C}{\mathrm{d}\tau} &= C\cdot \Theta^T - \Theta^* \cdot C \\ &\quad + \im \sum_m \mathcal{K}_m^T \cdot C \cdot \mathcal{K}_m^T,
\end{aligned}
\end{equation}
where the dot indicates matrix multiplication, and we introduced the matrix $\Theta_{ij} = \mathcal{H}_{ij} - \frac{\im}{2}\sum_m(\mathcal{K}^2_m)_{ij}$. $\Theta$ is generally non-Hermitian. Remarkably, Eq.~\eqref{eq:CM_EOM} is linear in $C$ and hence remains solvable~\cite{Zunkovic_2014, Barthel_2022}. The last term on the right-hand side plays the role of interactions. 

To solve Eq.~\eqref{eq:CM_EOM}, standard vectorization~\cite{PhysRevA.87.012108, Chaduteau_2025} defines a superoperator/matrix $\Chat$, of shape $N^2 \times N^2$, acting on vectorized correlation matrices $\ket{C}=\sum_{i,j}C_{ij}\ket{i}\otimes\ket{j}$, such that $\im\frac{\mathrm{d}}{\mathrm{d}\tau}\ket{C}:=\Chat \ket{C}$.
This mathematical object is the protagonist of our work; it is what permits analytical treatment in this interacting system. Eigenvectors of $\Chat$ correspond to states whose correlation matrix evolves as $C(\tau) = e^{-\im\lambda \tau}C(0)$. The corresponding eigenvalue $\lambda$ has an imaginary part which dictates the decay rate,  while the real part governs oscillations. Due to the interaction term, an initial Gaussian state will evolve into a mixture of Gaussian states, and Wick's theorem does not hold. The tractability of our model does not stem from Gaussianity; it follows instead from the closure of equations of motion~\cite{Zunkovic_2014, Barthel_2022}. Thus, solvability coexists with genuinely non-Gaussian, interacting dynamics. 
\begin{figure*}[tp]
    \centering
    \includegraphics[width=2.05\columnwidth]{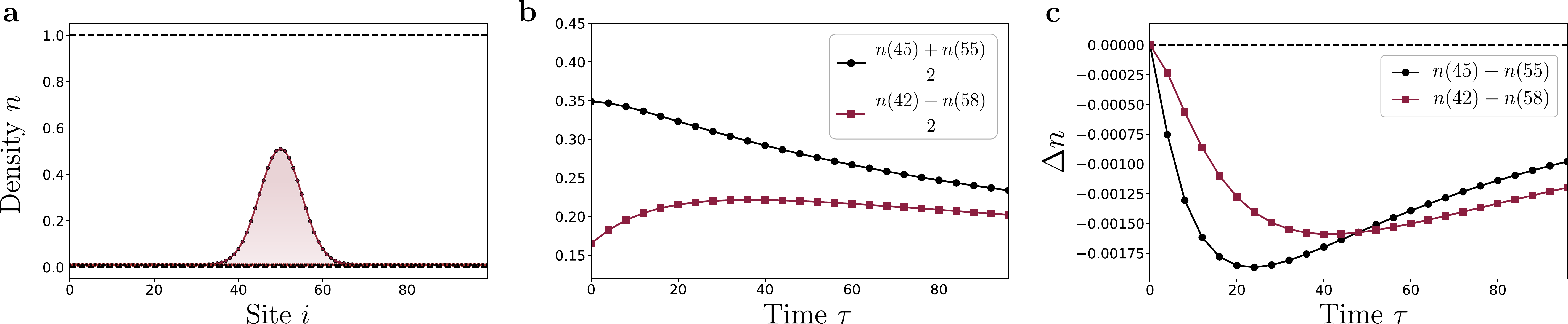}
    \caption{\small{\textbf{Robust asymmetric diffusion.} \textbf{a.} At time $\tau=0$, we perturb the steady state's correlation matrix with a diagonal matrix that forms a Gaussian profile, with center at the middle of the chain ($N=100$ here). We then numerically time-evolve according to Eq.~\eqref{eq:CM_EOM}.
    \textbf{b.} Corresponding time evolution for average densities located symmetrically from the middle. These relax back toward the same steady-state value. \textbf{c.} Time evolution of the difference between the same densities. We observe asymmetric diffusion, which is a consequence of the decohered band. The large characteristic timescale matches the $\Chat$ point gap being connected to the steady state. The scale of asymmetric diffusion does not diminish with increasing $N$~\cite{Note1}.}}
    \label{fig:dynamics_figure}
\end{figure*}

Consider first the eigenspectrum of $\Chat$ for our model under PBC, shown in Fig.~\ref{fig:model_spectrum}(c) for specific parameters. The large spectral region at lower imaginary values forms the continuous spectrum, with $\mathcal{O}(N^2)$ eigenvalues. This part derives from the first two terms on the right-hand side of Eq.~\eqref{eq:CM_EOM}. The central focus of this work is the other set of $\mathcal{O}(N)$ eigenvalues attached to the origin, commonly referred to as the discrete part of the spectrum. It is enabled by the interaction term in Eq.~\eqref{eq:CM_EOM} and, by analogy with interacting Hamiltonians, consists of ``bound-state" solutions. The discrete spectrum holds dynamical consequences for perturbations from the steady state. At short time scales, an initial state can couple to any eigenstate. For long times though, the state can only couple to discrete-spectrum eigenstates, since they have characteristically lower decay rates [see Fig.~\ref{fig:model_spectrum}(c)]. Thus, the spatial profile of these eigenstates dictates long-time dynamics. 


\emph{Decoherence-induced topology---}
We now study both the continuous and discrete parts of the $\Chat$ spectrum. Firstly, the former cannot have a point gap, namely a point in the complex plane around which the spectrum winds without crossing~\cite{kawabata_38_fold_way, PhysRevLett.124.086801}. Letting $\lambda_i$ be eigenvalues of $\Theta$, the continuum's eigenvalues are $\lambda_i-\lambda^*_j$ [see Eq.~\eqref{eq:CM_EOM}], where $i$ and $j$ are independent spatial indices.
The spectrum of $\Theta$ for our model, as shown in Fig.~\ref{fig:model_spectrum}(b), forms an ellipse in PBC. Then, in the thermodynamic limit $N \to \infty$, the continuous spectrum fills an entire disk~\cite{PhysRevB.105.165137}, matching Fig.~\ref{fig:model_spectrum}(c).

Non-trivial topology must thus stem from the discrete spectrum. Its structure can be explained most simply via symmetries of $\Chat$. For this, we write with explicit indices $\im (\mathrm{d}{C}_{ij}/
\mathrm{d}\tau)=\sum_{l,n}\mathcal{C}_{i,j,l,n}C_{ln}$, where $i,j,l,n$ are all spatial indices. Firstly, $\Chat$ enjoys a ``SWAP" symmetry $\SWAP$ that flips the input/output (or alternatively, ket/bra) indices, $\SWAP (A\otimes B)\SWAP^{-1}=B \otimes A$. Under this operation, $\mathcal{C}_{i,j,l,n}=-\mathcal{C}^*_{j,i,n,l}$. As a matrix expression, this gives $\SWAP \Chat \SWAP^{-1} = -\Chat^*$ [see Methods], identifying $\SWAP$ as a chiral symmetry.  Secondly, in PBC, $\Chat$ is invariant under simultaneous lattice translations of all four indices, $\mathcal{C}_{i,j,l,n}=\mathcal{C}_{i+1,j+1,l+1,n+1}$. This is a weak symmetry in the sense of Refs.~\onlinecite{Buca_2012, PhysRevX.6.041031}: the interaction term in Eq.~\eqref{eq:CM_EOM} locks ket and bra translations together, precluding a conserved momentum while permitting block diagonalization of $\Chat$. Each block is then labeled by a ``momentum" $k$~\footnote{When postselecting, two separate translational symmetries arise, giving two momentum labels to $\Chat$. This corresponds to a strong translational symmetry~\cite{Buca_2012}.}. With this, $\Chat$ is fully determined by the momentum-space expression (derived in the Methods)
\begin{equation}\label{eq:C_ab_k_simple_expression}
    \Cabk := \sum_{i-l}\mathcal{C}_{i,j,l,n}e^{\im k (i-l)},
\end{equation}
where $a=j-i$, $b=n-l$, and $k$ is in the Brillouin zone $\frac{2\pi}{N}\mathbb{Z}$. Unlike usual tight-binding problems, the size of the Bloch matrix $\Cabk$ scales with system size $N$, implying infinitely many eigenvalues at each $k$. We discuss this further in the Methods. The SWAP symmetry in momentum space gives
\begin{equation}\label{eq:momentum_C_ab_K_mymmetry}
\mathcal{C}_{ab}(k) = - \mathcal{C}^*_{-a,-b}(-k)e^{\im k (b-a)}.
\end{equation}
Defining $U_{ab}(k)=\delta_{a,-b}e^{\im k b}$, we see that $U(k)\mathcal{C}(k)U^\dagger(k)=-\mathcal{C}^*(-k)$. Thus, the spectrum is symmetric under $\lambda(k) \rightarrow-\lambda(-k)^*$. Eigenvalues either come in pairs or are fixed points of this symmetry.

To see the implications of these symmetries on the discrete spectrum, consider first the regime $A \neq 0$ and $B=0$ in Eq.~\eqref{eq:interacting_model}. For $t=0$, sites are dynamically decoupled and subject to on-site dephasing. In the occupancy number basis, a density matrix's off-diagonal elements are exponentially suppressed. For small $t$, coherent terms only cause slight dispersion. Long-lived modes then consist of classical mixtures. Standard Green's function methods~\cite{economou2006green, Note1} on $\Cabk$ give a new band of eigenvalues $\lambda(k) = -\im A^2 + \sqrt{\left[2t(1-\cos k)\right]^2-A^4}.$ This forms the discrete spectrum, lying on the imaginary axis, with width dictated by $t^2/A^2$ (obtained from expanding the square root~\cite{Note1}). As this spectrum consists of near-classical modes, we also refer to it as the ``decohered" band. It includes the steady state at $k=0$. In addition, we have inversion symmetry $\mathcal{C}_{ab}(k)=\mathcal{C}_{-a,-b}(-k)$ [see SM~\cite{Note1}], meaning the spectrum is symmetric under $\lambda(k)\rightarrow \lambda(-k)$. Combining with the SWAP symmetry gives a $\lambda(k) \rightarrow -\lambda^*(k)$ symmetry. As there can only be one discrete eigenvalue for each $k$, perturbations preserving inversion symmetry must give purely imaginary perturbed eigenvalues. Thus, there is no point gap here.

\begin{figure*}[tp]
    \centering
    \includegraphics[width=1.8\columnwidth]{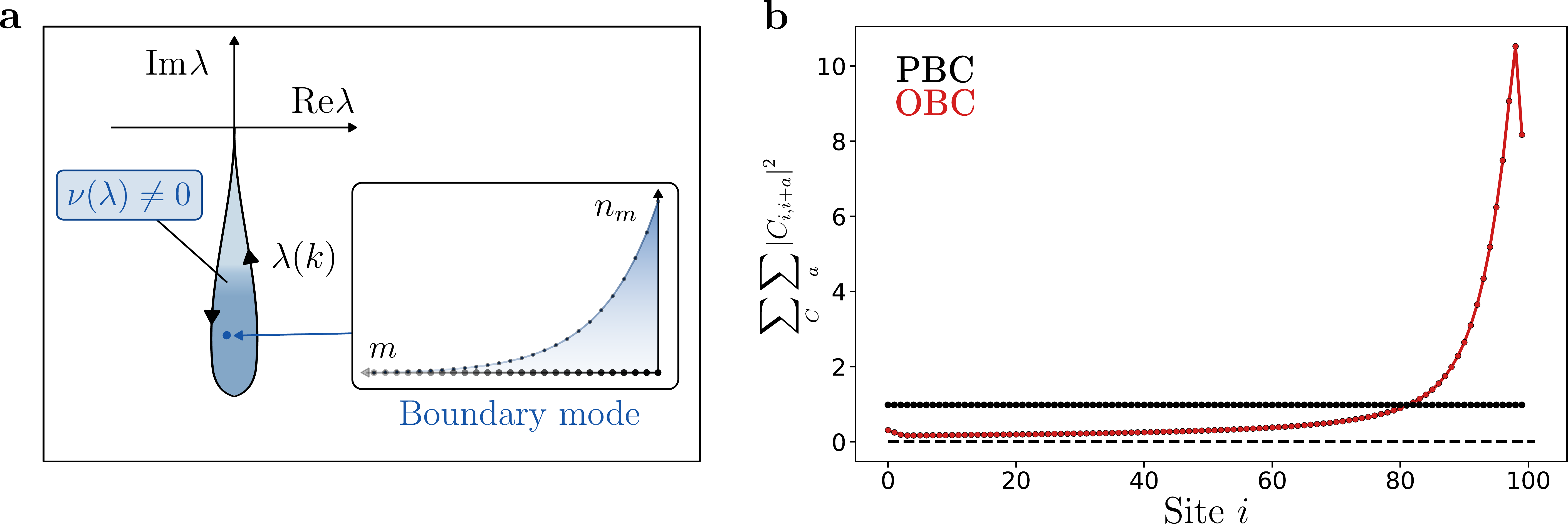}
    \caption{\small{\textbf{Decoherence-induced skin effect.} \textbf{a.} Schematic of the PBC point gap (black) of the $\Chat$ spectrum for our model [see Eq.~\eqref{eq:interacting_model}]. The spectrum winds against $k$ in the complex plane. In semi-infinite boundary conditions (i.e. with one edge), the bulk-boundary correspondence~\cite{PhysRevLett.124.086801} guarantees the existence of a ``boundary" mode (blue point), with density localized at the edge, at any point $\lambda \in \mathbb{C}$ around which the PBC winding $\nu(\lambda) \neq 0$. Decohered states are most localized at the bottom of the point gap, as reflected by the color gradient. \textbf{b.} Localization $\sum_C\sum_a |C_{i,i+a}|^2$ over all decohered eigenmodes $C$ of $\Chat$, in PBC vs OBC. We used the same parameters as in Fig.~\ref{fig:model_spectrum}. While the profile is uniform in PBC, there is clear localization in OBC. The height of the OBC peak scales linearly with $N$. The smaller peak on the left side does not scale and is a finite-size effect. When the decohered band reverses winding direction, the OBC localization occurs on the opposite edge~\cite{Note1}.}}
    \label{fig:skin_effect_figure}
\end{figure*}
Turning on $B$ breaks inversion symmetry, so decohered-band eigenvalues can develop nonzero real parts. The SWAP symmetry still pairs up eigenvalues $\lambda(k), \lambda(-k)$ with opposite real parts and equal imaginary parts. By periodicity in $k$, this band \textit{must} open a point gap. This proves the presence of directional winding, as observed in the spectrum in Fig.~\ref{fig:model_spectrum}(c) and (d). The dynamical consequences are shown in Fig.~\ref{fig:dynamics_figure}. Perturbing the steady state correlation matrix with a Gaussian density profile and time-evolving with $\Chat$, we observe asymmetric diffusion, whose direction is dictated by the winding of the decohered band point gap. Crucially, the nontrivial topology implies robustness of this effect to perturbations.

The presence of this point gap is sharply diagnosed by a winding number over the whole PBC spectrum
\begin{equation}\label{eq:winding_number_definition}
    \nu(\lambda) = \frac{1}{2\pi \im} \int_{0}^{2\pi}\mathrm{d}k \frac{\partial}{\partial k}\log \det \left[\mathcal{C}(k) - \lambda\id\right],
\end{equation}
where we wrote $\mathcal{C}(k)$ as a matrix in the indices $a,b$ [see Eq.~\eqref{eq:C_ab_k_simple_expression}]. Here $\lambda \in \mathbb{C}$.
In the Methods, we argue that the bulk-boundary correspondence still holds here. In an infinite system with a single edge~\cite{PhysRevLett.124.086801}, whenever $\nu(\lambda) \neq 0$, a localized eigenmode of $\Chat$ at eigenvalue $\lambda$ will exist, as shown in Fig.~\ref{fig:skin_effect_figure}(a).

Beyond this specific model, the chiral symmetry $\SWAP$ places $\Chat$ in class AI of the 38-fold way of non-Hermitian matrices~\cite{kawabata_38_fold_way}. This class's point gap classification in 1D is $\mathbb{Z}$, diagnosed by a winding number. We expect arbitrary integer windings to be realizable in the discrete spectrum via longer-range hoppings in $H$ and $K_m$, with winding number in Eq.~\eqref{eq:winding_number_definition}.

\emph{Many-body Lindbladian skin effect---}
The decohered eigenstates of $\Chat$ are localized, as we analytically show in the SM~\cite{Note1}. We ansatz a solution to the bulk recurrence relation of the form $C_{ij}=\phi_{j-i} z^{i+j}$, for $z$ and $\phi_a$ some complex coefficients. The localization behavior is not apparent from simply looking at e.g. the diagonal elements of Eq.~\eqref{eq:CM_EOM}. It only appears when keeping terms $C_{i,i+a}$ for $a \leq 2$. Letting $z=e^\kappa$ and expanding in $\kappa$ near the steady state, we find leading-order solutions with scaling
\begin{equation}\label{eq:analytical_localization}
\begin{aligned}
    \mathrm{Re}(\kappa) &\propto (t^2-g^2)|\lambda|, \\ \mathrm{Im}(\kappa) &\propto \sqrt{|\lambda|},
\end{aligned}
\end{equation}
with $g=AB/2$ and scaling coefficients that depend on $t,A,B$ and are defined in the SM~\cite{Note1}.
The localization length $\xi = 1/|\mathrm{Re}(\kappa)|$ scales as $1/|\lambda|$, which is in marked contrast with the Hatano-Nelson model, where it is constant. In particular, $\xi$ diverges as $\lambda \to 0$: the steady state $\rho_S \propto \id$ is delocalized, as expected. We verify the above scalings numerically in the SM~\cite{Note1}. Boundary conditions then dictate how to appropriately superpose these solutions. In Fig.~\ref{fig:skin_effect_figure}(b), we confirm this localization by plotting $\sum_C\sum_a |C_{i,i+a}|^2$, where the first sum $\sum_C$ is over all decohered-band eigenstates $C$, against site $i$. This profile is completely uniform in PBC, but localized at the right edge of the chain in OBC.

This result establishes the asymmetric diffusion of Fig.~\ref{fig:dynamics_figure} in OBC as well: long time dynamics are dictated by the decohered band, whose point gap yields a skin effect.
Moreover, as Eq.~\eqref{eq:analytical_localization} suggests, we numerically find a divergence of $\xi$ at $t=\pm g$. The PBC point gap closes at these points, in accordance with the bulk-boundary correspondence. Remarkably though, the point gap in $\Theta$ [see Fig.~\ref{fig:model_spectrum}(b)] remains open throughout. To our knowledge, this represents the first case of a decoherence-induced topological phase transition in a many-body Lindbladian system. For the $|t|>|g|$ case in PBC, the decohered band winds oppositely to $\Theta$ [see Fig.~\ref{fig:skin_effect_figure} and SM~\cite{Note1}]. This implies that the continuous spectrum (generated by $\Theta$) has a skin effect on the opposite side of the chain as the decohered point gap. On the other hand, for $|t|<|g|$, both the decohered band and $\Theta$ wind in the same direction; their eigenmodes have densities localized at the same edge. Thus, for fixed $t$ and $A$, simply increasing $B$ can cause a topological phase transition, reversing the winding number of $\Chat$. 

\emph{Discussion---}
We have introduced a lattice model of environment-induced decoherence, whose Lindbladian realizes many-body point-gap topology. The topology in our model resides in the near-classical sector of the dynamics: as in quantum Brownian motion~\cite{CALDEIRA1983587, PhysRevLett.46.211}, the long-lived decohered states are near-classical mixtures in the occupancy basis, with coherences mostly dephased; the continuous spectrum consists of shorter-lived ``quantum'' states involving coherent hopping. While the familiar single-particle skin effect involves the latter, our many-body skin effect relies on the former. This model thus realizes topology not in the coherent dynamics of a system, but in the stochastic relaxation through which it loses coherence, constituting the first instance of genuinely decoherence-induced topology in an open quantum system. The trivial steady state~\cite{PhysRevA.87.012108} confirms that the effect is purely dynamical. This sharply distinguishes our work from prior numerical work on Lindbladian skin effects which arise from non-Hermitian jumps~\cite{PhysRevB.108.155114, PhysRevLett.127.070402, Eisler_2011, PhysRevLett.123.080601}; here, the jumps are Hermitian, the mechanism is correlated density-current noise, and the model is solved analytically.

Our work suggests several natural extensions. Of immediate interest is identifying how to amplify the decohered point gap, causing larger robust asymmetric diffusion. More broadly, Lindbladians built from non-commuting Hermitian jumps may generically host interaction-induced topology; the general mechanism behind this is yet to be established. Since our model physically occurs in the Markovian/high-temperature limit, it is worth seeking a low-temperature completion and investigating non-Markovian effects. Also, for $B \neq 0$ and $A=0$, our model realizes the quantum symmetric simple exclusion process~\cite{Eisler_2011, Znidaric_2010_1, PhysRevLett.123.080601}. 
It would be interesting to explore consequences of our work on such processes. Lastly, our decoherence-induced topology occurs in the noise-averaged dynamics, opening the door for viable experimental realizations. The density and current jumps are well-within reach for ultracold atoms via laser-assisted hopping with spontaneous emission~\cite{PhysRevA.82.063605,Dalibard_1985,PhysRevLett.127.070402, PhysRevLett.123.080601}, and also in arrays of superconducting qubits~\cite{Naeij2025, Shen2025}.

\paragraph{Acknowledgments:}
We thank S. Chirame and A. Romito for insightful discussions. A.C. acknowledges support from Imperial College London via a President’s PhD Scholarship. This work was supported by a UKRI Future Leaders Fellowship MR/Y017331/1. A.P. is supported by the U.S. Department of Energy, Office of Science, Office of Advanced Scientific Computing Research, through the Exploratory Research for Extreme Scale Science (EXPRESS) program under Award No. DE-SC0026216.

\bibliography{refs}

\clearpage
\section*{Methods}
\subsection*{Reconstructing the Lindbladian spectrum}
Generalizing discussions in the main text, given an arbitrary Hermitian Hamiltonian $H$ and arbitrary quantum jumps $K_m$, the Lindblad equation is (in our imaginary factor convention)
\begin{equation}\label{eq:lindbladian_general}
\begin{aligned}
\im \dot{\rho} = \mathcal{L}\rho &= \left[H,\rho\right] - \frac{\im}{2}\sum_m\{K^\dagger_m K_m, \rho\} \\ &\quad + \im \sum_m K_m\rho K^\dagger_m.
\end{aligned}
\end{equation}
Restricting to Hermitian $K_m$, we obtain a double-commutator as in Eq.~\eqref{eq:lindbladian_quadratic_hermitian_jumps} of the main text. For quadratic, $U(1)$-conserving $H$ and $K_m$, one can check closure of equations of motion~\cite{Zunkovic_2014, Barthel_2022, Eisler_2011, PhysRevA.87.012108} via the adjoint action of the Lindbladian. Firstly, a few lines of algebra (detailed in the SM~\cite{Note1}) will show
\begin{equation}\label{eq:Ldag_action_single_particle_generalr}
\begin{aligned}
    \mathcal{L}^\dagger c_i &= -\sum_j \Theta_{ij}c_j, \\
     \mathcal{L}^\dagger c^\dagger_i &= -\sum_j c^\dagger_j(\Theta^\dagger)_{ji}.
\end{aligned}
\end{equation}
Thus, single-particle dynamics are closed and determined by $\Theta$ [see Eq.~\eqref{eq:CM_EOM}]. In other words, left eigenvectors of $\Theta$ give left eigenvectors of the Lindbladian. 

Next, two-particle observables are determined by the correlation matrix, for which more algebra results in Eq.~\eqref{eq:CM_EOM} of the main text [see SM~\cite{Note1}]. We can reconstruct a left eigenvector $X_C$ of $\mathcal{L}$ given a left eigenvector $\ket{C} = \sum_{i,j}C_{ij}\ket{i}\otimes \ket{j}$ of $\Chat$. The latter satisfies $\Chat^\dagger \ket{C} = \lambda \ket{C}$. Explicitly, let $X_C = \frac{1}{2^{N-2}}\sum_{ij}C_{ji}\cdag_i c_j$, which satisfies
\begin{equation}\label{eq:X_C_correlation_matrix}
\begin{aligned}
    \mathrm{Tr}(X_C \cdag_ic_j) &= \sum_{lk}C_{kl}(\delta_{ik}\delta_{lj}+\delta_{ij}\delta_{kl})  \\ &= C_{ij} + \delta_{ij}\mathrm{Tr}(C).
\end{aligned}
\end{equation}
Using Eq.~\eqref{eq:lindbladian_quadratic_hermitian_jumps} and~\eqref{eq:CM_EOM} [see SM for details~\cite{Note1}], we find
\begin{equation}
\begin{aligned}
    \mathcal{L}^\dagger X_C &= \frac{1}{2^{N-2}}\sum_{ijl}\left[\Theta^*_{li}C_{jl}-C_{li}\Theta^T_{jl}\right]\cdag_i c_j \\ &-\frac{1}{2^{N-2}}\im\sum_m (\mathcal{K}_m^T)_{li}C_{ji}(K^T_m)_{jk}\cdag_ic_j \\
    &= \frac{1}{2^{N-2}}\sum_{ijl}[\Chat^\dagger \ket{C}]_{ji}\cdag_ic_j \\
    &= \lambda X_C.
\end{aligned}
\end{equation}
Thus, two-particle observables, relevant for skin effects, are entirely determined by $\Chat$. This dramatically simplifies the initially exponential complexity of diagonalizing the full Lindbladian to $\mathcal{O}(N^2)$ complexity. 
\subsection*{Momentum-space expressions}
By translational invariance of $H$ and of the $K_m$, $\Theta$ is also translationally invariant, i.e. $\Theta_{i,j}=\Theta_{i+1,j+1}$. For the correlation matrix,
$\im\ket{\dot{C}}_{ij}=\sum_{l,n}\mathcal{C}_{i,j,l,n}\ket{C}_{ln}$ directly requires
\begin{equation}\label{eq:C_superoperator_explicit}
\begin{aligned}
    \mathcal{C}_{i,j,l,n} &= \delta_{il}\Theta_{jn}-\Theta^*_{il}\delta_{jn} \\ &\quad +\im\sum_m (\mathcal{K}_m)_{li}(\mathcal{K}_m)_{jn}.
\end{aligned}
\end{equation}
Using the vectorization map $X_{ijln}=A_{il} B_{jn} \rightarrow \hat{X} = B \otimes A$, the superoperator $\Chat$ takes the form
\begin{equation}
\begin{aligned}
\Chat
&=
\Theta \otimes \id
-
\id \otimes \Theta^*
\\
&\quad
+
\im \sum_m \mathcal{K}_{m}\otimes \mathcal{K}_m^T .
\end{aligned}
\end{equation}
where $\id$ represents the identity operator and $\otimes$ denotes a Kronecker product of matrices. Then, we have $\mathcal{C}_{i,j,l,n}=\mathcal{C}_{i+1,j+1,l+1,n+1}$. We let $a=j-i$ and $b=n-l$ so that $\mathcal{C}_{i,j,l,n} = \mathcal{C}_{i,i+a,l,l+b}$. Translational invariance allows shifting by $-i$ to obtain $\mathcal{C}_{i,j,l,n}=\mathcal{C}_{0,a,l-i,l-i+b}$, which now only depends on three indices $a,b$, and $l-i$.
The dynamical equation can be rewritten as
\begin{equation}
\im\dot{C}_{i,i+a}=\sum_{l,b}\mathcal{C}_{i,i+a,l,l+b}C_{l,l+b}.
\end{equation}
We now introduce $k \in \frac{2\pi}{N}\mathbb{Z}$; multiplying both sides by $e^{\im k i}$ and summing over $i$ gives
\begin{equation}
\begin{aligned}
\im\sum_i \dot{C}_{i,i+a}e^{\im k i}
&=
\sum_{i,l,b}
\Bigl[
\mathcal{C}_{0,a,l-i,l-i+b}
e^{\im k(i-l)}
\\
&\qquad\qquad\times
C_{l,l+b}e^{\im k l}
\Bigr].
\end{aligned}
\end{equation}
We define the left-hand side as $\im\sqrt{N}\dot{C}_{k,a}$. We also shift the $i$ sum at fixed $l$ to a sum over $i-l$ to obtain
\begin{equation}
\im\dot{C}_{k,a}=\sum_{b}\Cabk C_{k,b},
\end{equation}
where (omitting the comma between indices $a$ and $b$)
\begin{equation}
    \Cabk := \sum_{i-l}\mathcal{C}_{0,a,l-i,l-i+b}e^{\im k (i-l)}.
\end{equation}
This is the equivalent of the Bloch momentum-space matrix for tight-binding Hamiltonians. Plugging in the explicit form Eq.~\eqref{eq:C_superoperator_explicit} will give
\begin{equation}\label{eq:C_ab_k_expression}
 \begin{aligned}
 \Cabk &= \Theta_{ab}-\Theta_{ba}^*e^{\im k(b-a)} \\ &+ \im\sum_{i-l,m} (\mathcal{K}_m)_{li}(\mathcal{K}_m)_{l+b,i+a}^* e^{\im k (i-l)}.
 \end{aligned}
\end{equation}
The first two terms on the right-hand side form the continuous/postselected part of the spectrum; the last term is the interaction term, and arises from quantum jumps. We obtain explicit expressions for our model in the SM~\cite{Note1}.

The SWAP symmetry states that $\mathcal{C}_{i,j,l,n}=-\mathcal{C}^*_{j,i,n,l}$, as can be explicitly checked on Eq.~\eqref{eq:C_superoperator_explicit}. Then, we have
\begin{equation}
\begin{aligned}
    \mathcal{C}^*_{ab}(k) &=\sum_{i-l}\mathcal{C}^*_{i,j,l,n}e^{\im k (l-i)} \\
    &= -\sum_{i-l} \mathcal{C}_{j,i,n,l}e^{\im k (l-i)} 
    \\ 
    &= -\sum_{l-i} \mathcal{C}_{0,-a,n-j,-b+n-j}e^{\im k (n-j+a-b)}, 
\end{aligned}
\end{equation}
where we used translational invariance to shift indices by $-j$. The sum index is $i-l$; when fixing $a,b$, this corresponds to summing over $j-n$. Thus
\begin{equation}
\begin{aligned}
\mathcal{C}^*_{ab}(k)e^{\im k (b-a)}
    &= -\sum_{j-n} \mathcal{C}_{0,-a,n-j,-b+n-j}e^{\im k (n-j)} \\
    &= -\mathcal{C}_{-a,-b}(-k), 
\end{aligned}
\end{equation}
matching Eq.~\eqref{eq:momentum_C_ab_K_mymmetry}. This can be explicitly checked on Eq.~\eqref{eq:C_ab_k_expression} using translational invariance of $\Theta$.
\subsection*{Bulk-boundary correspondence}
At a given $E \in \mathbb{C}$, consider the Hermitian matrix built from the $N^2\times N^2$ matrix $\Chat$ as
\begin{equation}
    \tilde{\mathcal{C}} = \begin{pmatrix}
        0 & \Chat-E \\
        \Chat^\dagger - E^* & 0
    \end{pmatrix}.
\end{equation}
Here $0$ denotes a null matrix of the same size as $\Chat$. Specializing to our case of study, for any $E$ inside the point gap of $\Chat$, $\tilde{\mathcal{C}}$ is gapped. It also automatically satisfies the chiral symmetry $\Gamma \tilde{\mathcal{C}} \Gamma^{-1}= -\tilde{\mathcal{C}}$, where $\Gamma = \sigma_z$. This places $\tilde{\mathcal{C}}$ in class AIII of the tenfold way for Hermitian matrices. As $\tilde{\mathcal{C}}$ is gapped, its eigenvectors decay exponentially in real space. Explicitly, given a threshold $\epsilon > 0$, there exists a $d \in \mathbb{N}$ such that $|C_{k,a}| < \epsilon$ for all $|a|\geq d$. This $d$ will typically be of order of the correlation length of the state. Thus, as far as decohered states are concerned, we can truncate the Bloch matrix $\mathcal{C}_{ab}(k)=0$ for all $|a|,|b| \geq d$. This is not true for eigenstates in the continuum of bands constructed from $\Theta$.

With this truncation, the decohered state eigenproblem reduces to a simple SSH-type problem with a finite-size unit cell. The bulk-edge correspondence of non-Hermitian Hamiltonians~\cite{PhysRevLett.124.086801} then carries over. There is the additional subtlety that the indices $a,b$ of $\mathcal{C}_{i,i+a,l,l+b}$ are relative, meaning they satisfy $a \geq -i, b \geq -l$. As these appear in our Bloch matrix $\Cabk$, we have a unit cell size that changes with position on the lattice. The above truncation however stabilizes the unit cell size to a value determined by $d$ in the bulk. The question is analogous to asking if the bulk-boundary correspondence persists if we remove some degrees of freedom near the boundary. Treating this edge deletion itself as a perturbation, as long as the bulk gap is not closed in $\tilde{\mathcal{C}}$, the bulk-boundary correspondence continues to hold.

In vectorized form, eigenstates of $\Chat$ take the form $\ket{C}=\sum_{i,j}C_{i,i+a}\ket{i}\otimes\ket{i+a}$. The skin effect in OBC implies exponential decay of the band diagonals $C_{i,i+a}$ at fixed relative index $a$, as in Fig.~\ref{fig:skin_effect_figure}(b).
\end{document}


\title{Supplementary Material for\\ ``Topology from Decoherence"}

\author{Alexandre Chaduteau}
\affiliation{Blackett Laboratory, Imperial College London, London SW7 2AZ, United Kingdom}

\author{Derek K. K. Lee}
\affiliation{Blackett Laboratory, Imperial College London, London SW7 2AZ, United Kingdom}

\author{Frank Schindler}
\affiliation{Blackett Laboratory, Imperial College London, London SW7 2AZ, United Kingdom}

\author{Abhinav Prem}
\affiliation{Physics Program, Bard College, 30 Campus Road, Annandale-on-Hudson, New York 12504, USA}

\maketitle

\tableofcontents

\section{Lindbladians with quadratic jump operators}
Consider a density operator $\rho$ describing the state of a quantum system. For example, if the system is in a pure state described by $|\psi\rangle$, then $\rho = |\psi\rangle \langle \psi|$. With this definition, the Lindblad equation~\cite{lindblad1976generators, gorini1976completely, Daley04032014, breuer2002theory} is a dynamical equation for the density matrix that can be derived under a series of assumptions, the most prominent being the Born approximation, Markovianity, and the rotating-wave approximation~\cite{breuer2002theory}. It can be expressed as
\begin{equation}\label{eq:lindblad_dynamics}
\mathrm{i} \frac{\mathrm{d} \rho}{\mathrm{d} \tau} := \mathcal{L}(\rho) = \left[ H, \rho \right] + \mathrm{i} \sum_m \left[ {K_m \rho K^\dagger_m - \frac{1}{2}\left\{ K^\dagger_m K_m, \rho \right\} } \right],
\end{equation}
where $H$ is a Hermitian Hamiltonian and $K_m$ are “jump operators", which are operators on Fock space that describe interactions between the system and the environment. For our purposes, studying one-dimensional systems (with $N$ lattice sites), these jumps are labelled by an integer $m$, and there can be any number\footnote{Although in the model we present here, we have as many jumps as lattice sites, i.e. $m \in \{1, \ldots N\}$.} of such jumps~\cite{breuer2002theory}. When all jumps are zero, we recover Schrödinger/von Neumann dynamics for closed quantum systems. Note that we adopt the convention whereby decay frequencies are directly incorporated in the definition of the jump operators.

In the case of quadratic Hermitian jump operators $K_m=K^\dagger_m$, the extra dissipative term becomes
\begin{equation}\label{eq:lindbladian_K_part}
    \mathcal{L}_K(\rho) := \im\sum_m \left(K_m\rho \Kdag_m - \frac{1}{2}\{\Kdag_m K_m, \rho\}\right) = -\frac{\im}{2}\sum_m\left[K_m, \left[K_m,\rho\right]\right].
\end{equation}
With these quadratic jump operators, an initial Gaussian state will become non-Gaussian under time evolution. As in the main text, we study fermions $c_i, \cdag_i$ obeying canonical anti-commutation relations $\{\cdag_i,c_j\}=\delta_{ij}, \; \{c_i,c_j\}=0$. Let us consider the case of particle-number-conserving Hamiltonian and Hermitian quadratic jumps
\begin{equation}
\begin{aligned}
    H &= \sum_{i,j=1}^N \mathcal{H}_{ij}\cdag_i c_j, \\
    K_m &= \sum_{i,j=1}^N(\mathcal{K}_m)_{ij}\cdag_ic_j,
\end{aligned}
\end{equation}
where the matrices $\mathcal{H}_{ij}$ and $(\mathcal{K}_m)_{ij}$ are all Hermitian in $i,j$. One can obtain the equation of motion for the correlation matrix $C_{ij}=\langle \cdag_i c_j \rangle$ via $\frac{\mathrm{d}C_{ij}}{\mathrm{d}\tau}=\mathrm{Tr}(\dot{\rho}\cdag_i c_j) = -\im \mathrm{Tr}(\mathcal{L}(\rho) \cdag_i c_j)$ as
\begin{equation}\label{eq:CM_EOM}
\begin{aligned}
    \frac{\mathrm{d}C}{\mathrm{d}\tau} &= \im\Theta^* \cdot C - C\cdot \im \Theta^T + \sum_m \mathcal{K}_m^T \cdot C \cdot \mathcal{K}_m^T,
\end{aligned}
\end{equation}
where we defined the $N\times N$ matrix
\begin{equation}
    \Theta_{ij} := \mathcal{H}_{ij} - \frac{\im}{2}\sum_m (\mathcal{K}_m^2)_{ij}.
\end{equation}
In Eq.~\eqref{eq:CM_EOM}, the first and second terms (which are Hermitian conjugates of each other) are responsible for coherent hopping in the chain. The new term $\sum_m \mathcal{K}_m^T\cdot C \cdot \mathcal{K}_m^T$ can be seen as an additional ``interaction" term.

The equation of motion for the correlation matrix [Eq.~\eqref{eq:CM_EOM}] itself takes the form of a Lindbladian. Vectorizing $C = \sum_{ij}C_{ij}\ket{i}\bra{j}$ to $\ket{C} = \sum_{ij}C_{ij}\ket{i}\otimes \ket{j}$ gives
\begin{equation}\label{eq:vectorised_CM_EOM}
    \im\ket{\dot{C}} := \Chat \ket{C} = \left[ \Theta \otimes \id - \id \otimes \Theta^* + \im\sum_m \mathcal{K}_m \otimes \mathcal{K}^T_m\right]\ket{C}.
\end{equation}
The quadratic terms, some of which are inside $\Theta$, can altogether be written as $-\frac{\im}{2}\sum_m(\mathcal{K}_m \otimes \id - \id \otimes \mathcal{K}_m^T)^2$. 
Per the main text, $\Chat$ is a $N^2 \times N^2$ matrix. Assuming it is invertible, Eq.~\eqref{eq:vectorised_CM_EOM}
is formally solved by
\begin{equation}
    \ket{C(\tau)} = e^{-\im\mathcal{\Chat}\tau}  \ket{C(0)}.
\end{equation}
The steady state's correlation matrix corresponds to $\ket{\dot{C}} = 0$ and is proportional to the identity in our case~\cite{PhysRevA.87.012108, Barthel_2022}. Generically, $\Chat$ is non-Hermitian and can be non-diagonalizable.
Its size is quadratic in $N$, allowing to solve for dynamics of quadratic observables for much larger systems than by explicit diagonalisation of $\mathcal{L}$.
\subsection{Mapping $\mathcal{C}$ eigenmodes to $\mathcal{L}$ eigenmodes}\label{subsec:mapping_C_to_L}
We first construct single-particle eigenmodes of $\mathcal{L}$ from those of $\Theta$. From the adjoint action of the Lindbladian, one can obtain
\begin{equation}\label{eq:Ldag_action_single_particle_generalr}
\begin{aligned}
    \mathcal{L}^\dagger c_i &= -\sum_j \Theta_{ij}c_j, \\
     \mathcal{L}^\dagger c^\dagger_i &= -\sum_j c^\dagger_j(\Theta^\dagger)_{ji}.
\end{aligned}
\end{equation}
If we enforce translational invariance via $(\mathcal{K}_m)_{ij}=(\mathcal{K}_{m+1})_{i+1,j+1}$, then $\sum_m(\mathcal{K}^2_m)_{i+1,j+1}=\sum_{u=m-1}(\mathcal{K}^2_u)_{i,j} = \sum_m(\mathcal{K}^2_m)_{i,j}$. Assuming translational invariance of the Hamiltonian via $\mathcal{H}_{i+1,j+1}=\mathcal{H}_{ij}$, we have $\Theta_{ij} = \Theta_{i+1,j+1}$ as well. 

Consider a left eigenvector $\ket{L_a}$ of $\Theta$, with components written $L^a_i$. Then the operator $X_a = \sum_i L^a_i c_i$ (and its dagger counterpart) is a right eigenmode of $\mathcal{L}^\dagger$, i.e. a left eigenmode of the Lindbladian. This reconstructs $2N$ of the $4^N$ eigenmodes of the Lindbladian. Next, let us establish the relation between eigenvectors of $\Chat$ and those of $\mathcal{L}$.
Using Eq.~\eqref{eq:lindbladian_K_part}, by methods in Ref.~\onlinecite{Chaduteau_2025}, we obtain
\begin{equation}
    \mathcal{L}_K^\dagger Y = +\frac{\im}{2}\sum_m \left[K_m,\left[K_m, Y\right]\right].
\end{equation}
Using 
\begin{equation}
    \left[\cdag_ic_j, \cdag_kc_l\right] = \delta_{jk}\cdag_ic_l - \delta_{il}\cdag_kc_j
\end{equation}
twice, we obtain
\begin{equation}
    \mathcal{L}^\dagger_K \cdag_i c_j = \frac{\im}{2}\sum_{ml}\left[(\mathcal{K}_m^2)_{li}\cdag_lc_j + (\mathcal{K}_m^2)_{jl}\cdag_i c_l\right] - \im \sum_{mlk}(\mathcal{K}_m)_{li}(\mathcal{K}_m)_{jk}\cdag_l c_k.
\end{equation}
Then the total adjoint action is
\begin{equation}\label{eq:Ldag_action_cdag_ic_j_general}
\begin{aligned}
\mathcal{L}^\dagger\cdag_ic_j &= \sum_l\left[\left(\mathcal{H}^*_{il} + \frac{\im}{2}\sum_m(\mathcal{K}_m^2)_{li}\right)\cdag_lc_j + \left(-\mathcal{H}_{jl}+\frac{\im}{2}\sum_m(\mathcal{K}_m^2)_{jl}\right)\cdag_ic_l\right]  \\ &-\im\sum_{mlk}(\mathcal{K}_m)_{li}(\mathcal{K}_m)_{jk}\cdag_lc_k \\
&= \sum_l\left[\Theta^*_{il}\cdag_lc_j - \cdag_ic_l\Theta^T_{lj} \right] -\im\sum_{mlk}(\mathcal{K}_m)^T_{il}\cdag_lc_k(\mathcal{K}_m)^T_{kj}.
\end{aligned}
\end{equation}
An alternative way of deriving Eq.~\eqref{eq:CM_EOM} starts from this action and uses the definition of the adjoint operator such that $ \mathrm{Tr}\left[\mathcal{L}(\rho) X^\dagger\right] = \mathrm{Tr}\left[(\mathcal{L}^\dagger X)^\dagger \rho\right]$ for all operators $X$.
Say the $\Theta$ eigenvector $\ket{L_a}$ has eigenvalue $T_a$. Then let $X_{ab} := \sum_{ij}L^{a*}_iL^b_j\cdag_ic_j$. We have
\begin{equation}
    \mathcal{L}^\dagger X_{ab} = (T^*_a-T_b) X_{ab} - \im\sum_{mlkij}(\mathcal{K}_m)_{li}(\mathcal{K}_m)_{jk}L^{a*}_iL^b_j\cdag_l c_k.
\end{equation}
Thus, single-particle left-eigenstates (those of $\Theta$) do not directly determine two-particle left-eigenstates due to the interaction term. Without it, we do reconstruct the postselected~\cite{breuer2002theory, Chaduteau_2025} spectrum of $\Chat$ completely, by which we mean the first two terms involving $\Theta$ and $\Theta^*$ in Eq.~\eqref{eq:vectorised_CM_EOM}. 

On the other hand, we can ask whether eigenvectors of $\Chat$ determine those of $\mathcal{L}$. Indeed, let $\rho_C$ such that $\mathcal{L}(\rho_C) = \lambda_C \rho_C$. Then $C_{ij}:=\mathrm{Tr}(\rho_C \cdag_i c_j)$ satisfies $\partial_\tau C_{ij} = -\im\mathrm{Tr}(\mathcal{L}(\rho_C) \cdag_i c_j) = -\im\lambda_C C_{ij}$. Vectorizing gives $\Chat \ket{C} = \lambda_C \ket{C}$. We can also reverse-engineer a left eigenvector $X_C$ of $\mathcal{L}$ given a left eigenvector of $\Chat$. For this, consider $\Chat^\dagger(C) = \lambda C$. Then let $X_C = \frac{1}{2^{N-2}}\sum_{ij}C_{ji}\cdag_i c_j$, which satisfies
\begin{equation}\label{eq:X_C_correlation_matrix}
    \mathrm{Tr}(X_C \cdag_ic_j) = \sum_{lk}C_{kl}(\delta_{ik}\delta_{lj}+\delta_{ij}\delta_{kl}) = C_{ij} + \delta_{ij}\mathrm{Tr}(C).
\end{equation}
Using Eq.s~\eqref{eq:Ldag_action_cdag_ic_j_general} and~\eqref{eq:CM_EOM} (and relabelling indices in the sums) we have
\begin{equation}
\begin{aligned}
    \mathcal{L}^\dagger X_C &= \frac{1}{2^{N-2}}\sum_{ijl}\left[\Theta^*_{li}C_{jl}-C_{li}\Theta^T_{jl}-\im\sum_m (\mathcal{K}_m^T)_{li}C_{ji}(K^T_m)_{jk}\right]\cdag_ic_j \\
    &= \frac{1}{2^{N-2}}\sum_{ijl}\Chat^\dagger(C)_{ji}\cdag_ic_j \\
    &= \lambda X_C.
\end{aligned}
\end{equation}
Thus, left eigenmodes of $\Chat$ correspond to left eigenmodes of $\mathcal{L}$. These correspond to the same correlation matrix, up to a factor proportional to the identity in Eq.~\eqref{eq:X_C_correlation_matrix}. This is the two-particle analog of the fact that $\Theta$ left eigenvectors map to single-particle left eigenmodes of $\mathcal{L}$.

\subsection{The Lindbladian skin effect remains a polynomial problem for quadratic jumps}
Consider an eigenmode $\rho_C$ of $\mathcal{L}$ with eigenvalue $\lambda_C$, so that under time evolution $\rho_C(\tau) = e^{-\im\lambda_C  \tau}\rho_C$. 
We have the corresponding equation $\dot C_{ij}(\tau) = \mathrm{Tr}(\dot\rho_C(\tau)\cdag_ic_j) = -\im\lambda_C \mathrm{Tr}(\rho_C(\tau)\cdag_ic_j)$ i.e. $\dot C(\tau) = -\im\lambda_C C(\tau)$. So $C$ is an eigenvector of $\Chat$ of eigenvalue $\lambda_C$. Yet the Lindbladian has $4^N$ eigenmodes but $\Chat$ has only $N^2$; how is this possible? Clearly for all of these, the eigen-equation $\dot{C} = \lambda C$ must be satisfied. There are then three possibilities: (1) the eigenvalue $\lambda$ is zero, meaning the corresponding $C$ is a steady state solution; (2) $C = 0$; (3) for degenerate eigenvalues $\lambda_C$, we can also have linearly dependent $C$, which is a specific constrained case that we do not consider here. Thus, \emph{for any Lindbladian whose steady state degeneracy does not scale exponentially with system size, and for which the correlation matrix equation of motion closes, we have at most $O(N^2)$ eigenmodes of $\mathcal{L}$ that have non-zero correlation matrix}. This directly implies that solving for skin effects is a polynomial ($\mathcal{O}(N^2)$) problem.

\section{Momentum space version of $\mathcal{C}$}\label{sec:momentum_space}
As explained in Sec.~\ref{subsec:mapping_C_to_L}, translational invariance of $\mathcal{H}$ and $\mathcal{K}_m$ implies that $\Theta$ is translationally invariant i.e. $\Theta_{i,j}=\Theta_{i+1,j+1}$, or in other words, $\Theta_{i,j}=\Theta_{i-j,0}$. Let us write the equation of motion for the correlation matrix as
$\im\ket{\dot{C}}_{ij}=\sum_{l,n}\mathcal{C}_{ij,ln}\ket{C}_{ln}$, which from Eq.~\eqref{eq:CM_EOM} requires
\begin{equation}\label{eq:C_superoperator_explicit}
    \mathcal{C}_{ij,ln}= \delta_{il}\Theta_{jn}-\Theta^*_{il}\delta_{jn}+\im\sum_m (\mathcal{K}_m)_{li}(\mathcal{K}_m)_{jn}.
\end{equation}
Using the vectorization map $X_{ijln}=A_{il} B_{jn} \rightarrow \hat{X} = B \otimes A$, the superoperator $\Chat$ takes the same form as before,
\begin{equation}
    \Chat = \Theta \otimes \id - \id \otimes \Theta^* + \im \sum_m \mathcal{K}_{m}\otimes \mathcal{K}^T_m,
\end{equation}
where $\id$ represents the identity operator and $\otimes$ denotes a Krönecker product of matrices.
Again, a few lines will show that $\mathcal{C}_{i,j,l,n}=\mathcal{C}_{i+1,j+1,l+1,n+1}$. This corresponds to saying that dynamics, which map a particle-hole pair's center of mass $(l+n)/2$ to another position $(i+j)/2$, does so regardless of the position on the lattice. 
The corresponding translation operator acts on the two-particle space via
\begin{equation}
 \hat{T}\ket{i,j}:=\ket{i+1,j+1},
\end{equation}
such that $\hat{T}\Chat \hat{T}^{-1}=\Chat$. We note that translational symmetry on ket indices only would give $\mathcal{C}_{i,j,l,n} = \mathcal{C}_{i+1,j+1,l,n}$; similarly, on bra indices only it would be $\mathcal{C}_{i,j,l,n} = \mathcal{C}_{i,j,l+1,n+1}$. The interaction term removes these as independent symmetries.

Now, let $a=j-i$ and $b=n-l$, so that $\mathcal{C}_{i,j,l,n} = \mathcal{C}_{i,i+a,l,l+b}$. Translational invariance allows shifting all indices by $-i$ to obtain $\mathcal{C}_{i,j,l,n}=\mathcal{C}_{0,a,l-i,l-i+b}$, which now only depends on three indices $a,b$, and $l-i$. The dynamical equation can be rewritten as
\begin{equation}
    \im\dot{C}_{i,i+a}=\sum_{l,b}\mathcal{C}_{i,i+a,l,l+b}C_{l,l+b}.
\end{equation}
We now introduce $k \in \frac{2\pi}{N}\mathbb{Z}$; multiplying both sides by $e^{\im k i}$ and summing over $k$ gives
\begin{equation}
    \im\sum_i \dot{C}_{i,i+a}e^{\im k i} = \sum_{i,l,b}\mathcal{C}_{0,a,l-i,l-i+b}e^{\im k(i-l)}C_{l,l+b}e^{\im k l}.
\end{equation}
We now define the left-hand side as $\im\sqrt{N}\dot{C}_{k,a}$. We also shift the $i$ sum at fixed $l$ to a sum over $i-l$ to obtain
\begin{equation}
\im\dot{C}_{k,a}=\sum_{b}\Cabk C_{k,b},
\end{equation}
where (omitting the comma between indices $a$ and $b$)
\begin{equation}\label{eq:Cab_k_definition}
    \Cabk := \sum_{i-l}\mathcal{C}_{0,a,l-i,l-i+b}e^{\im k (i-l)}.
\end{equation}
This is the equivalent of the Bloch momentum-space matrix for tight-binding Hamiltonians. We can also rewrite the superoperator as
\begin{equation}\label{eq:general_C_superoper_double_ket}
\begin{aligned}
    \Chat &= \sum_{i,j,l,n}\mathcal{C}_{ijln}(\ket{i}\bra{l})\otimes(\ket{j}\bra{n}) \\
    &= \sum_{i,l,a,b}\mathcal{C}_{0,a,l-i,l-i+b}\ket{i,i+a}\bra{l,l+b} \\
    &= \frac{1}{N}\sum_{i,l,a,b,k}\Cabk e^{\im k (l-i)}\ket{i,i+a}\bra{l,l+b}.
\end{aligned}
\end{equation}
We now define $\ket{k,a} = \frac{1}{\sqrt{N}}\sum_i e^{-\im k i} \ket{i,i+a}.$ These are eigenvectors of $\hat{T}$ as $\hat{T} \ket{k,a} = e^{\im k }\ket{k,a}$. By the usual route to the momentum-space Bloch Hamiltonian, we obtain
\begin{equation}
    \begin{aligned}
        \Chat &= \sum_{a,b, k}\mathcal{C}_{ab}(k)\ket{k,a}\bra{k,b}.
    \end{aligned}
\end{equation}
Plugging in the explicit form in Eq.~\eqref{eq:C_superoperator_explicit}, we obtain
\begin{equation}\label{eq:Cab_k_expression}
\begin{aligned}
\Cabk = \Theta_{ab}-\Theta_{ba}^*e^{\im k(b-a)} + \im\sum_{i-l}\sum_m (\mathcal{K}_m)_{li}(\mathcal{K}_m)_{l+b,i+a}^* e^{\im k (i-l)}.
\end{aligned}
\end{equation}
One issue is that this Bloch matrix's size scales with system size $N$. A similar case arises in Ref.~\cite{PhysRevX.15.021016}; in their case, assuming finite-range jumps, one can truncate this matrix to a finite size, independent of $N$. This is not true in our case due to the terms involving $\id$. In other words, the continuous part of the spectrum has infinitely many eigenvalues at each $k$ in the thermodynamic limit. On the other hand, the discrete part of the spectrum, which is where point-gap topology can occur [see main text], has finitely many eigenvalues at each $k$. The postselected part of $\Chat$, which generates the continuum, has a second translational invariance in $b-a$. This corresponds to independent translational symmetries on ket/bra layers. As a result, the postelected part actually has a second momentum label $\kappa$. Explicitly, we write
\begin{equation}\label{eq:Cpostk_explicit}
    \Cpostk = \Theta_{ab}-\Theta^*_{ba}e^{\im k (b-a)}.
\end{equation}
Since $\Theta$ is translationally invariant, we have the second translational symmetry $\Tauhat$ such that
\begin{equation}
    \Tauhat \Chat \Tauhat^{-1} = \Chat,
\end{equation}
associated with translational invariance of kets/bras independently. Letting $\ket{k,\kappa}=\frac{1}{\sqrt{N}}\sum_a e^{\im \kappa a}\ket{k,a}$, a similar procedure to the above yields
\begin{equation}
    \hat{\mathcal{C}}^{\mathrm{post}} = \sum_{k,\kappa} \mathcal{C}^{\mathrm{post}}(k,\kappa)\ket{k,\kappa}\bra{k,\kappa},
\end{equation}
where using $d:=b-a$
\begin{equation}\label{eq:double_momentum_C_definition}
    \mathcal{C}^{\mathrm{post}}(k,\kappa) := \sum_d \mathcal{C}^{\mathrm{post}}_d(k) e^{\im \kappa d}.
\end{equation}
Clearly then, 
\begin{equation}
    \mathcal{C}_\mathrm{post}(k,\kappa) = \Theta(k+\kappa) - \Theta(\kappa)^*.
\end{equation}
Setting $\kappa=0$ allows us to retrieve the winding of $\Theta(k)$.
As for the previous truncation problem, we see that the part that cannot be truncated is precisely $\Cpostk$, meaning the total $\mathcal{C}_{ab}(k)$ matrix has infinitely many identical blocks that repeat [see also Ref.~\cite{PhysRevX.15.021016}].

The SWAP symmetry of $\Chat$ induces a symmetry on the Bloch matrix $\mathcal{C}_{ab}(k)$. This symmetry states that $\im\mathcal{C}_{ij,ln}=-\im\mathcal{C}^*_{ji,nl}$, which can be explicitly checked on Eq.~\eqref{eq:C_superoperator_explicit}. Then, we have
\begin{equation}
\begin{aligned}
    \mathcal{C}^*_{ab}(k) &=\sum_{i-l}\mathcal{C}^*_{i,j,l,n}e^{\im k (l-i)} \\
    &= -\sum_{i-l} \mathcal{C}_{j,i,n,l}e^{\im k (l-i)} 
    \\ 
    &= -\sum_{l-i} \mathcal{C}_{0,-a,n-j,-b+n-j}e^{\im k (n-j+a-b)}, 
\end{aligned}
\end{equation}
where we used translational invariance to shift indices by $-j$. The sum index is $i-l$; when fixing $a,b$, this corresponds to summing over $j-n$. Thus
\begin{equation}
\begin{aligned}
\mathcal{C}^*_{ab}(k)e^{\im k (b-a)}
    &= -\sum_{j-n} \mathcal{C}_{0,-a,n-j,-b+n-j}e^{\im k (n-j)} \\
    &= -\mathcal{C}_{-a,-b}(-k), 
\end{aligned}
\end{equation}
as in the main text.
\section{Decoherence-induced skin effect}\label{sec:spectral_NHSE}
We now consider the following model:
\begin{equation}\label{eq:interacting_model}
    \begin{aligned}
        H &= t\sum_{m=1}^N (\cdag_{m+1}c_m + \text{h.c.}), \\
        K_m &= A\cdag_mc_m+\im B (\cdag_{m}c_{m+1} - \cdag_{m+1}c_{m}),
    \end{aligned}
\end{equation}
where $t,A,B$ are all real parameters. Periodic boundary conditions (PBC) are implemented via $c_1 \equiv c_{N+1}$. The $B$ terms are the nearest-neighbor current operator, and the integer $m$ runs from $1$ to $N$. The open-boundary counterpart of this model uses the last jump $K_N=A\cdag_N c_N$. 
\subsection{Explicit matrices}
In PBC, we have
\begin{equation}
    \sum_{m=1}^N \mathcal{K}_m = \begin{pmatrix}
        A & \im B & 0 \ldots \\
        -\im B  & A & \im B \ldots \\
        \ldots & \ddots & \cdots
    \end{pmatrix},
\end{equation}
which is translationally invariant.
In addition,
\begin{equation}
    \sum_m \mathcal{K}_m^2 = \begin{pmatrix}
        A^2 + 2B^2 & \im AB & 0 \ldots \\
        -\im AB  & A^2 + 2B^2 & \im AB \ldots \\
        \ldots & \ddots & \cdots
    \end{pmatrix},
\end{equation}
which is a nearest-neighbour hopping matrix. We thus have a a point-gapped $\Theta$, since
\begin{equation}
    \Theta = \mathcal{H} - \frac{\im}{2}\sum_m \mathcal{K}^2_m = \sum_m \left[\left(t+\frac{1}{2}AB\right)\ket{m}\bra{m+1}+\left(t-\frac{1}{2}AB\right)\ket{m+1}\bra{m}\right] - \frac{\im}{2}(A^2+2B^2)\id,
\end{equation}
and its Fourier components are $\Theta(k) = 2t\cos k - \frac{\im}{2}(A^2+ 2B^2 + 2AB\sin k)$.
The corresponding matrix $\mathcal{C}_{ab}(k)$ [see Eq.~\eqref{eq:Cab_k_expression}] is a tight-binding/Toeplitz PBC matrix. The postselected part is
\begin{equation}\label{eq:Cpostk_explicit_model}
    \Cpostk = \begin{pmatrix}
        -\im(A^2+2B^2) & (t+g)-(t-g)e^{\im k} & 0 &\ldots & (t-g)-(t+g)e^{-\im k} \\
        (t-g)-(t+g)e^{-\im k} & -\im(A^2+2B^2) & (t+g)-(t-g)e^{\im k} & \ldots & 0 \\
        0 & (t-g)-(t+g)e^{-\im k} & \ldots & \ldots & \ldots \\
        \vdots \\
        (t+g)-(t-g)e^{\im k} & \ldots
    \end{pmatrix},
\end{equation}
where $g = \frac{1}{2}AB$. Its eigenvectors can be solved via the regular Bloch-wave ansatz, which will give the continuum's dispersion relation in $k$ and a second momentum label $\kappa$.
As for the mixing term $\im\CQk := \Cabk - \Cpostk$, we find
\begin{equation}\label{eq:CQk_explicit}
    \CQk = \begin{pmatrix}
        A^2+2B^2\cos k & \im AB & 0 &\ldots & -\im ABe^{-\im k} \\
        -\im AB & 0 & 0 & \ldots & -B^2e^{-\im k} \\
        0 & 0 & 0 & \ldots & 0\\
        \vdots & \ldots & \ddots & \cdots & \vdots \\
        \im ABe^{\im k} & -B^2e^{\im k} & 0 & \ldots & 0
    \end{pmatrix},
\end{equation}
which clearly does not have the additional translational symmetry of the postselected part [see Sec.~\ref{sec:momentum_space}].

The point gap in $\Theta$ implies a non-Hermitian skin effect of single-particle left eigenmodes $\rho_a = \sum_i L^a_i c_i$ (and its dagger counterpart) for $L^a$ a left eigenmode of $\Theta$. This does not directly correspond to a skin effect in the two-point observables $\langle \cdag_i c_j\rangle$ due to the mixing term of $\Chat$. When turning off the mixing term, the decohered point gap of $\Chat$ disappears, implying that it is a purely ``jump-induced" effect; it disappears upon postselection.

\begin{figure}
    \centering
    \includegraphics[width=0.75\linewidth]{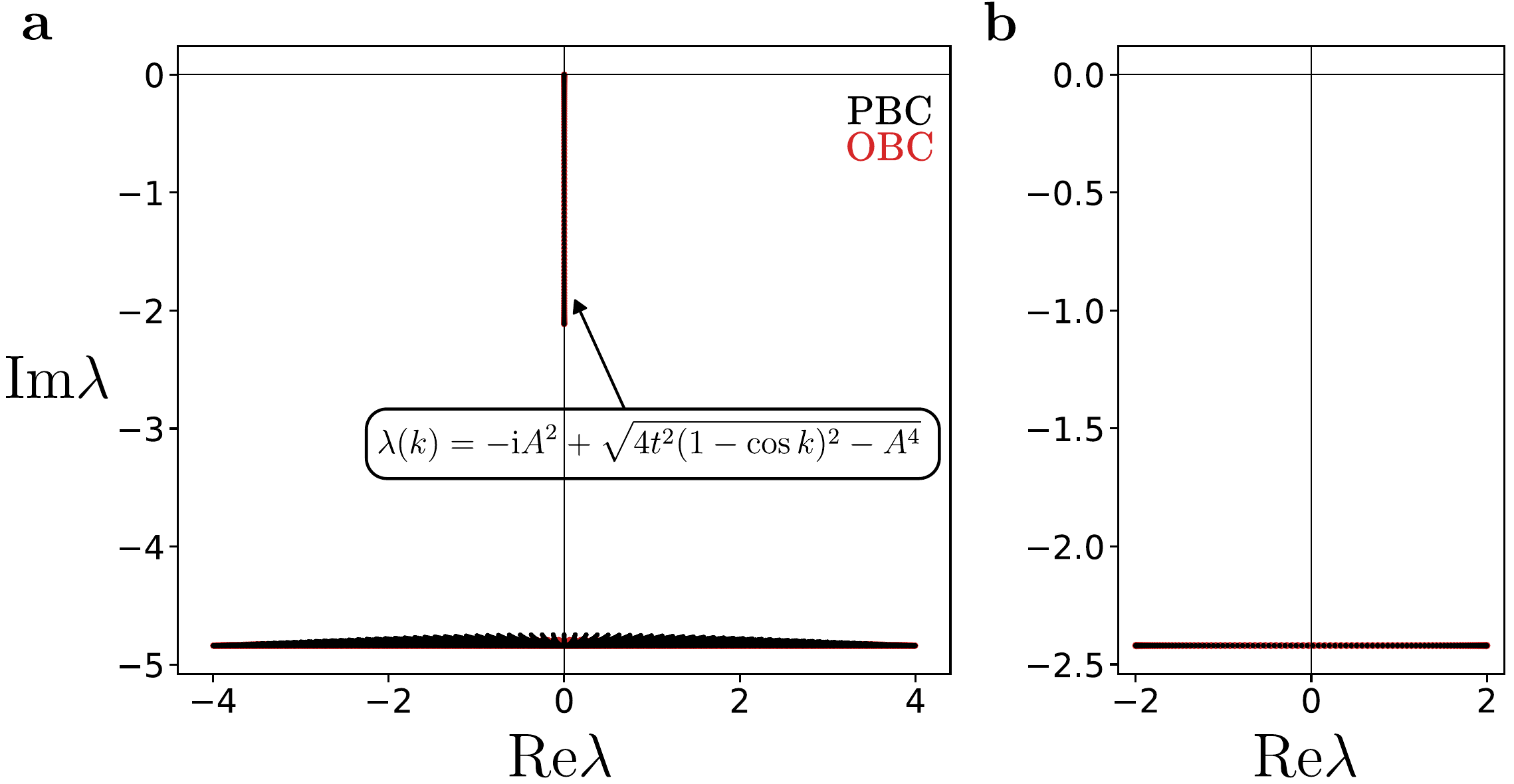}
    \caption{Spectrum of the model without current measurements. \textbf{a.} Spectrum of $\Chat$ for $t=1, A=2.2, B=0$. When $B=0$, the model obeys inversion symmetry [Eq.~\eqref{eq:C_inversion_symmetry}] and the point gap is closed. Focusing on decohered-state eigenvalues, these match exactly the ones derived via Green's functions [Sec.~\ref{subsec:spectrum_analysis}]. 
    \textbf{b.} Spectrum of $\Theta$, for which the point gap is closed as well, as expected from its hoppings involving $t \pm \frac{AB}{2}$.}
    \label{fig:model_spectrum_B_0}
\end{figure}
\subsection{Explaining the spectrum}\label{subsec:spectrum_analysis}
Let us explain aspects of the spectrum [Fig.~1 of the main text]. Firstly, the mirror symmetry about the imaginary axis is due to the fact that eigenmodes come in conjugate pairs: $\mathcal{L}(\rho) = \lambda \rho \implies \mathcal{L}\rho^\dagger=-\lambda^* \rho^\dagger.$
These eigenmodes are not Hermitian if their eigenvalue is not purely imaginary, and they are also traceless by trace-preservation. Secondly, the continuum is understood simply from the two momentum labels of $\Cpostk$ [see Eq.~\eqref{eq:Cpostk_explicit}]; expressing its solutions as Bloch waves reproduces the same spectrum. Thirdly, for $B=0$, our model admits inversion symmetry, which maps site $m$ to site $-m$ up to an affine shift defining the inversion center. Ignoring that affine term, the corresponding symmetry is $\mathcal{C}_{i,j,l,n}=-\mathcal{C}_{-i,-j,-l,-n}.$ Using Eq.~\eqref{eq:Cab_k_definition}, this implies
\begin{equation}\label{eq:C_inversion_symmetry}
\begin{aligned}
    \mathcal{C}_{ab}(k) &= \sum_{i-l}\mathcal{C}_{0,-a,i-l,i-l-b}e^{\im k (i-l)} \\
    &= \sum_{l-i}\mathcal{C}_{0,-a,i-l,i-l-b}e^{-\im k (l-i)} \\ 
    &= \mathcal{C}_{-a,-b}(-k).
\end{aligned}
\end{equation}
Then, if $\lambda(k)$ is an eigenvalue, so is $\lambda(-k)$. Combining this with the SWAP symmetry implies pairs $\{\lambda(k),-\lambda^*(k)\}$. The continuum indeed comes in such pairs, at opposite real values and equal imaginary values. For decohered states, there can only be a single eigenvalue at momentum $k$. This forces $\lambda(k)=-\lambda^*(k)$ i.e. purely imaginary decohered-state eigenvalues.

Let us focus on the decohered band. The perturbation in $\CQk$ for $B=0$ is simple enough to allow analytical treatment. This can be treated as a bound state problem, akin to the Cooper instability problem~\cite{economou2006green}. In 1D, the density of states diverges near the bottom of the energy band, so there is a bound/decohered state for any nonzero perturbation strength. Assuming non-Hermiticity does not change the reasoning in Chapter 6 of Ref.~\cite{economou2006green}, Green's function methods analytically give the decohered-state eigenvalues as
\begin{equation}
    \lambda(k)=-\im A^2 \pm \sqrt{4t^2(1-\cos k)^2 - A^4}.
\end{equation}
The Green's function method used is nonperturbative and gives a solution to all orders, so it is also true for $A>t$, as in our case. Expanding the above in $t/A$, we have
\begin{equation}
\begin{aligned}
    \lambda(k) &=-\im A^2 \pm \im A^2\sqrt{1-\frac{4t^2(1-\cos k)^2}{A^4}} \\
    &= -\im A^2 \pm \im A^2 \left[1-\frac{2t^2(1-\cos k)^2}{A^4}\right].
\end{aligned}
\end{equation}
From this we see two branches, one near the origin and one near $-2\im A^2$, the latter of which we discard. These eigenvalues are purely imaginary, as expected from inversion symmetry. The decohered band width scales with $t^2/A^2$. We plot an example of this spectrum in Fig.~\ref{fig:model_spectrum_B_0}.

Crucially, perturbing with $B$ breaks the inversion symmetry of Eq.~\eqref{eq:C_inversion_symmetry}. Decohered-state eigenvalues are no longer required to satisfy $\lambda(k)=-\lambda^*(-k)$; in particular they can have a non-zero real part. This opens up the point gap, which is symmetric in $k$, by the SWAP symmetry of $\Chat$.

\subsection{Analytical solutions to the eigenvalue equation}
One can explicitly solve $\Chat \ket{C} = \lambda \ket{C}$ [see Eq.~\eqref{eq:vectorised_CM_EOM}] for the model Eq.~\eqref{eq:interacting_model}. Let us define $n_i = C_{ii}$, $l_i = C_{i,i+1}$, $r_i=C_{i+1,i}$. It is easiest to do it from the equation of motion rather than vectorizing; we obtain
\begin{equation}
\label{eq:locbulk}
    \begin{aligned}
        (\lambda + 2\im c - \im A^2)n_i &= \im B^2 (n_{i+1}+n_{i-1}) + (t-g)(r_{i-1}-l_{i-1})+(t-g)(l_i-r_i), \\
        (\lambda + 2\im c)r_i &= -(t+g)n_i + (t+g)n_{i+1} - \im B^2 l_i - (t+g)C_{i+2,i}+(t-g)C_{i+1,i-1}, \\
        (\lambda + 2\im c)l_i &= (t+g)n_i - (t+g)n_{i+1}-\im B^2 r_i +(t+g)C_{i,i+2}-(t-g)C_{i-1,i+1}, \\
        (\lambda+2\im c)C_{i,i+2} &= (t+g)\left[C_{i,i+3}-C_{i+1,i+2}\right] + (t-g)\left[l_i-C_{i-1,i+2}\right], \\
        (\lambda+2\im c)C_{i+2,i} &= (t+g)\left[C_{i+2,i+1}-C_{i+3,i}\right]+(t-g)\left[C_{i+2,i-1}-r_i\right], \\
        \ldots
    \end{aligned}
\end{equation}
for all bulk $i=2,\ldots N-1$. Boundary equations can similarly be obtained. However, we are interested in the localization scaling, so it suffices to solve these bulk equations. Boundary conditions will simply dictate how to appropriately superpose the solutions.

Only the density ($n_i$) and currents (involving $r_i$ and $l_i$) have terms coming from interactions. All $C_{ij}$ can be non-zero, though we expect their magnitude to decay with the relative index $j-i=a$. Let us now study solutions under different truncation orders.

\emph{Zeroth order truncation---}
To zeroth order, one can truncate by setting $C_{i,i+a}=0 \;\forall |a|\geq 1$. This gives an imaginary reciprocal tight-binding hopping in the $n_i$, so there is no skin effect there. This matches numerics, where the most classical states (the ones nearest to the steady state in the decohered band) have the least localization (and they are essentially most similar to the $B=0$ case).

\emph{First order truncation---}
To first order, we truncate via $C_{i,i+a}=0 \;\forall |a|\geq 2$. Let us also define $\Lambda = \lambda + 2\im c$. Then equations~\eqref{eq:locbulk} reduce to
\begin{equation}
    \begin{aligned}
        (\Lambda - \im A^2)n_i &= \im B^2 (n_{i+1}+n_{i-1}) + (t-g)(r_{i-1}-l_{i-1}+l_i-r_i), \\
        \Lambda r_i &= (t+g)(n_{i+1}-n_i) - \im B^2 l_i, \\
        \Lambda l_i &= -(t+g)(n_{i+1}-n_i)-\im B^2 r_i.
    \end{aligned}
\end{equation}
We now ansatz $C_{ij}=\phi_{j-i}z^{i+j}$, for $\phi_{j-i}$ and $z$ complex numbers. Then, in particular, $n_i = \phi_0 z^{2i}, \, l_i = \phi_1 z^{2i+1}, \, r_i = \phi_{-1}z^{2i+1}$. Plugging these into the above, we obtain
\begin{equation}
    \begin{aligned}
        \left[\Lambda - \im A^2-\im B^2(z^2+z^{-2})\right]\phi_0 &= \phi_1(t-g)(z-z^{-1}) - \phi_{-1}(t-g)(z-z^{-1}), \\
        \Lambda \phi_{-1}z &= (t+g)\phi_0(z^2-1) - \im B^2 \phi_1 z, \\
        \Lambda \phi_1 z &= -(t+g)\phi_0(z^2-1)-\im B^2 \phi_{-1}z.
    \end{aligned}
\end{equation}
Summing the last two equations gives
\begin{equation}
    (\Lambda + \im B^2)(\phi_1 + \phi_{-1})z=0,
\end{equation}
meaning either $\lambda + 2 \im c + \im B^2 = 0$, or $z=0$, or $\phi_1 = -\phi_{-1}$. Ignoring the first two cases, we obtain
\begin{equation}
    \phi_1 = \frac{1}{(\Lambda - \im B^2)z}(t+g)\phi_0 (1-z^2).
\end{equation}
Plugging this into the first equation, we obtain
\begin{equation}\label{eq:first_order_loc_eqn}
    \left[\Lambda - \im B^2\right]\left[\Lambda - \im A^2 - \im B^2(z^2+z^{-2})\right]=2(t^2-g^2)\left[2-(z^2+z^{-2})\right].
\end{equation}
This is a quadratic equation in the coefficient $z^2$. While the explicit solution for $z^2$ can be obtained, it is more illuminating to expand near the steady state which is completely delocalized. We let $z^2 = e^{\kappa}$. We have 
\begin{equation}
    \left[\lambda + \im (A^2+B^2)\right]\left[\lambda + \im B^2(2-(z^2+z^{-2}))\right]=2(t^2-g^2)\left[2-(z^2+z^{-2})\right].
\end{equation}
Expanding in $\kappa$ and rewriting $\lambda = -\im \Omega$, we obtain
\begin{equation}
    \Omega^2 - (A^2+B^2)\Omega + \Omega B^2 \kappa^2 - \left[(A^2+B^2)B^2+2(t^2-g^2)\right]\kappa^2 + \ldots = 0,
\end{equation}
where dots indicate terms of order $\kappa^4$. Let us consider a solution $\kappa = \im \kappa_1 \sqrt{\Omega}.$ Expanding in leading orders of $\Omega$ gives
\begin{equation}
    \Omega\left[-(A^2+B^2)+\Gamma\kappa^2_1\right] + \Omega^2\left[1-B^2\kappa^2_1\right] \approx 0,
\end{equation}
where we defined $\Gamma = (A^2+B^2)B^2 + 2(t^2-g^2)$. Neglecting $\Omega^2$, we get
\begin{equation}
    \kappa^2_1 = \frac{A^2+B^2}{\Gamma},
\end{equation}
which is real. This gives purely oscillatory solutions with no localization length. While one could try other solutions / expand to higher orders, we can actually treat the more general case of a second-order truncation.
\begin{figure}
    \centering
    \includegraphics[width=0.9\linewidth]{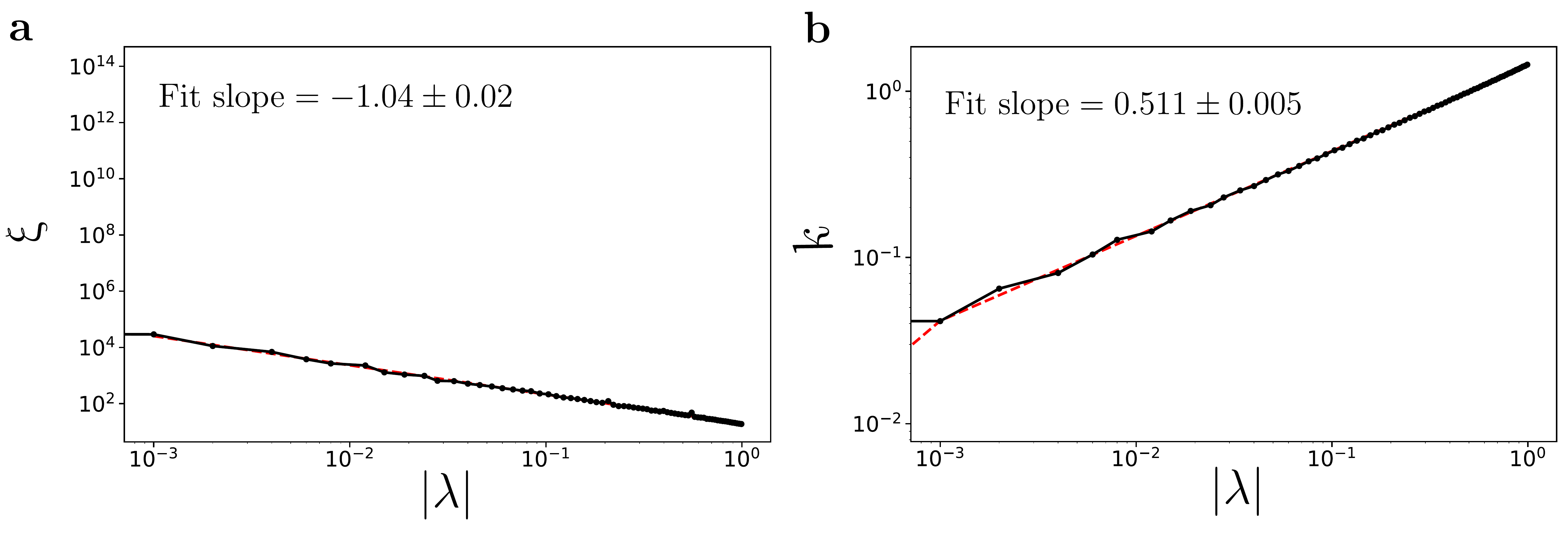}
    \caption{\textbf{} Numerical fits of the localization length $\xi$ (a) and the oscillation wavevector $k$ (b) of decohered eigenmodes near the steady state in OBC. Parameters are the same as in Fig.~1 of the main text, i.e. $t=1, A=2.2, B=0.5$. We fit the spatial profile of eigenmodes via the density $n_i$. The fit slopes roughly match the second-order predictions from Eq.~\eqref{eq:analytical_localization_scaling}.}
    \label{fig:numerical_localization_fits}
\end{figure}

\emph{Second order truncation---}
Let us now truncate to $C_{i,i+a}=0 \;\forall |a|\geq3$. With the same ansatz $C_{ij}=\phi_{j-i}z^{i+j}$, equations~\eqref{eq:locbulk} become
\begin{equation}\label{eq:locbulk_ansatz}
    \begin{aligned}
        \left[\Lambda-\im A^2 - \im B^2(z^2+z^{-2})\right]\phi_0 &= \phi_1(t-g)(z-z^{-1})-\phi_{-1}(t-g)(z-z^{-1}), \\
        \Lambda\phi_{-1}z &= \phi_0 (t+g)(z^2-1) - \im B^2 \phi_{1}z - \phi_{-2}\left[(t+g)z^2 - (t-g)\right], \\
        \Lambda\phi_1 z &= -\phi_0 (t+g)(z^2-1) - \im B^2 \phi_{-1}z + \phi_{2}\left[(t+g)z^2 - (t-g)\right], \\
        \Lambda\phi_{-2}z^2 &= \phi_{-1}z\left[(t+g)z^2-(t-g)\right], \\
        \Lambda\phi_2 z^2 &= -\phi_1 z\left[(t+g)z^2-(t-g)\right].
    \end{aligned}
\end{equation}
From this, defining $A_z = (t+g)z^2 - (t-g)$, we get
\begin{equation}
    \begin{aligned}
        \Lambda \phi_{-1}z &= \phi_0(t+g)(z^2-1) - \im B^2 \phi_1 z - \frac{\phi_{-1} z A^2_z}{\Lambda z^2}, \\
        \Lambda \phi_1 z &= -\phi_0(t+g)(z^2-1) - \im B^2 \phi_{-1}z - \frac{\phi_{1}z A^2_z}{\Lambda z^2}.
    \end{aligned}
\end{equation}
Computing their sum and difference gives, respectively,
\begin{equation}
\begin{aligned}
\Lambda(\phi_1+\phi_{-1})z &= -\im B^2(\phi_1+\phi_{-1})z - \frac{zA_z^2}{\Lambda z^2}(\phi_1+\phi_{-1}), \\
\Lambda (\phi_1 - \phi_{-1})z &= 2(t+g)(1-z^2)\phi_0 + \im B^2 (\phi_1 - \phi_{-1})z - (\phi_1 - \phi_{-1})\frac{zA^2_z}{\Lambda z^2}.
\end{aligned}
\end{equation}
From the first equation, either we have $\Lambda + \im B^2 + \frac{A^2_z}{\Lambda z^2}=0$, or $\phi_1 = -\phi_{-1}$, or $z=0$. From the second equation, we also have the case $\phi_1 = \phi_{-1}$, which requires either $t=-g$ (which we know numerically corresponds to a point gap closing) or $z=1$, implying no localization. Let us focus on the case $\phi_1 = -\phi_{-1}$. The difference equation yields
\begin{equation}
    \left[\Lambda -\im B^2+ \frac{A^2_z}{\Lambda z^2}\right]2\phi_1 z = 2\phi_0 (t+g)(1-z^2).
\end{equation}
\begin{figure}
    \centering
    \includegraphics[width=0.8\columnwidth]{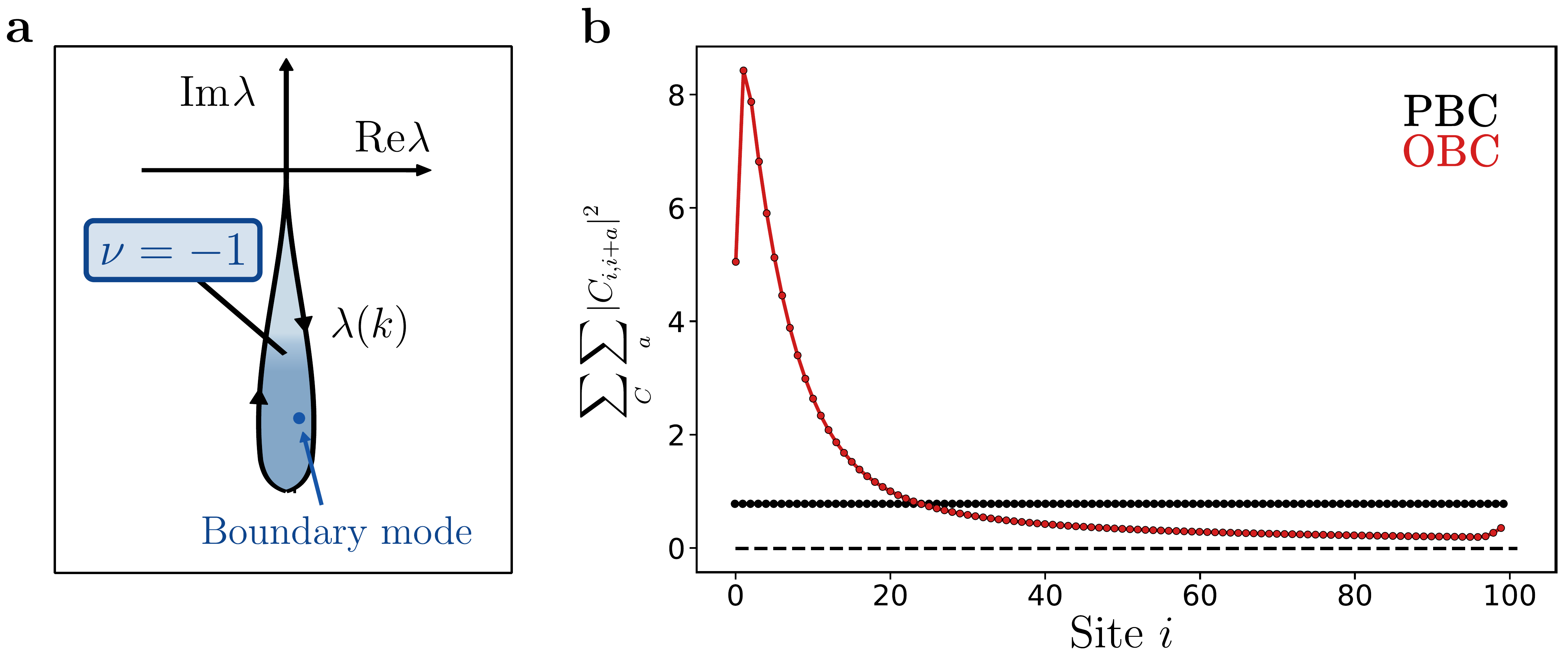}
    \caption{\textbf{a.} Schematic of the point gap with winding reversed compared to the case considered in the main text [see Fig. 3 there]. Parameters used here are $t=1,A=2.7,B=1.1, N=100$. The winding direction of the spectrum of $\Theta$ remains the same as in the main text. \textbf{b.} Corresponding localization plot of decohered states, as per Fig.~3 of the main text. The localization reverses edges here. As a result, the asymmetric diffusion reverses direction. Thus, increasing $B$ can cause a reversal of the winding direction of $\Chat$, while keeping that of $\Theta$ unchanged; this is an interaction-induced topological phase transition.}
    \label{fig:reversed_skin_effect}
\end{figure}
Plugging this into the first line of Eq.~\eqref{eq:locbulk_ansatz} gives 
\begin{equation}\label{eq:master_loclength}
    \left[\Lambda - \im B^2 + \frac{A^2_z}{\Lambda z^2}\right]\left[\Lambda - \im A^2 - \im B^2(z^2+z^{-2})\right] = 2(t^2-g^2)\left[2-(z^2+z^{-2})\right].
\end{equation}
This is the same equation as Eq.~\eqref{eq:first_order_loc_eqn} for the first-order truncation with an extra self-energy term. This no longer depends on any $\phi_a$ and allows us to derive a localization length. Again we rewrite $\lambda = -\im \Omega$ and let $z^2 = e^{\kappa}$. Expanding the two terms on the left-hand side in $\kappa$ separately gives
\begin{equation}
\begin{aligned}
    \left[\Lambda - \im A^2 - \im B^2(z^2+z^{-2})\right] &= -\im(\Omega + B^2 \kappa^2) + \ldots, \\
    \left[\Lambda - \im B^2 + \frac{A^2_z}{\Lambda z^2}\right] &= \im \left[\Gamma_0-(\gamma_1+\gamma^\prime_1)-(\gamma_1-\gamma^\prime_1)\kappa - \frac{\gamma_1+\gamma_1^\prime}{2}\kappa^2-\frac{\gamma_1-\gamma^\prime_1}{6}\kappa^3+\ldots\right],
\end{aligned}
\end{equation}
where the dots indicate higher orders of $\kappa$. We also defined $\Gamma_0 = A^2+B^2+2\frac{t^2-g^2}{A^2+2B^2}$, $\gamma_1 = \frac{(t+g)^2}{A^2+2B^2}$, $\gamma^\prime_1 = \frac{(t-g)^2}{A^2+2B^2}$. Let also $\Gamma_1 = \gamma_1-\gamma^\prime_1 = \frac{4tg}{A^2+2B^2}$, $\Gamma^\prime_1 = \gamma_1 + \gamma^\prime_1 = 2\frac{t^2+g^2}{A^2+2B^2}$. Plugging this into Eq.~\eqref{eq:master_loclength} gives the condition
\begin{equation}\label{eq:leading_order_scaling_equation}
    (\Gamma_0-\Gamma^\prime_1)\Omega - \Gamma_1 \Omega \kappa -\frac{\Gamma^\prime_1}{2}\Omega \kappa^2 + \Gamma_2\kappa^2 - B^2\Gamma_1\kappa^3 - \frac{\Gamma_1}{6}\Omega\kappa^3 +\ldots = 0,
\end{equation}
where we also defined $\Gamma_2 = (\Gamma_0-\Gamma^\prime_1) B^2 + 2(t^2-g^2)$.
We can then approximate the solution as
\begin{equation}
    \kappa \sim \kappa_0 \Omega \pm i \kappa_1 \sqrt{\Omega}.
\end{equation}
Plugging this into Eq.~\eqref{eq:leading_order_scaling_equation}, we get the consistency conditions (to leading order)
\begin{equation}
    \kappa_1^2 = \frac{\Gamma_0-\Gamma^\prime_1}{\Gamma_2} =\frac{A^2+B^2-\frac{4g^2}{A^2+2B^2}}{(A^2+B^2-\frac{4g^2}{A^2+2B^2})B^2+2(t^2-g^2)}
\end{equation}
and to next order
\begin{equation}
    \kappa_0 = \frac{\Gamma_1\Gamma_2 - B^2\Gamma_1(\Gamma_0-\Gamma^\prime_1)}{2\Gamma^2_2}=\frac{\Gamma_1(t^2-g^2)}{\Gamma^2_2}.
\end{equation}
Note here that we need $\Gamma_0 - \Gamma^\prime_1 > 0$ for this to make sense; this is always true for real $A,B$ since
\begin{equation}
\begin{aligned}
    \Gamma_0 - \Gamma^\prime_1 &= A^2 + B^2 + 2\frac{t^2-g^2}{A^2 + 2B^2} - 2\frac{t^2+g^2}{A^2+2B^2} \\
    &= \frac{A^4 + 2A^2 B^2 + 2 B^4}{A^2 + 2B^2} \geq 0.
\end{aligned}
\end{equation}
So the two relevant non-Bloch roots are
\begin{equation}\label{eq:analytical_localization_scaling}
\boxed{
    \kappa_{\pm} \approx \frac{\Gamma_1(t^2-g^2)}{\Gamma^2_2} \Omega \pm \im \sqrt{\frac{\Gamma_0-\Gamma^\prime_1}{\Gamma_2}} \sqrt{\Omega}}.
\end{equation}
The imaginary part gives the long-wavelength oscillation, while the common real part gives the exponential skin profile. From this, we can now read off the localization length
\begin{equation}
    \xi^{-1} = |\mathrm{Re}(\kappa)| = \left| \frac{\Gamma_1(t^2-g^2)}{\Gamma^2_2} \right| |\lambda|.
\end{equation}
This scaling matches a numerical fit, as shown in Fig.~\ref{fig:numerical_localization_fits}. In OBC, the eigenmodes lie on the imaginary axis, meaning their correlation matrix is Hermitian. This constrains the possible superpositions of the two solutions $\kappa_\pm$ here. As their real part is equal, fitting numerical eigenmodes to a single exponential does suffice to extract $\xi$. At $B = 0$, $\Gamma_1 = 0$ (since $g = AB/2$) and $\Gamma_2$ remains finite, so there is no localization, as expected. In addition, flipping the sign of $B$ flips the sign of $g$, which flips $\Gamma_1$, which dictates the side on which localization occurs. This also shows that the localization length diverges when $t = \pm g$. In that case, $\kappa_1^2=1$, so $\kappa \sim \im \sqrt{\Omega}$ still oscillates, which we verified numerically. Lastly, the fact that the localization plots [Fig.~\ref{fig:reversed_skin_effect} here and also Fig.~3 of the main text] peak not at the edge but slightly before is likely an effect of the boundary conditions, which we do not consider here.
\subsection{Dynamical consequences}
\begin{figure}
    \centering
    \includegraphics[width=0.5\linewidth]{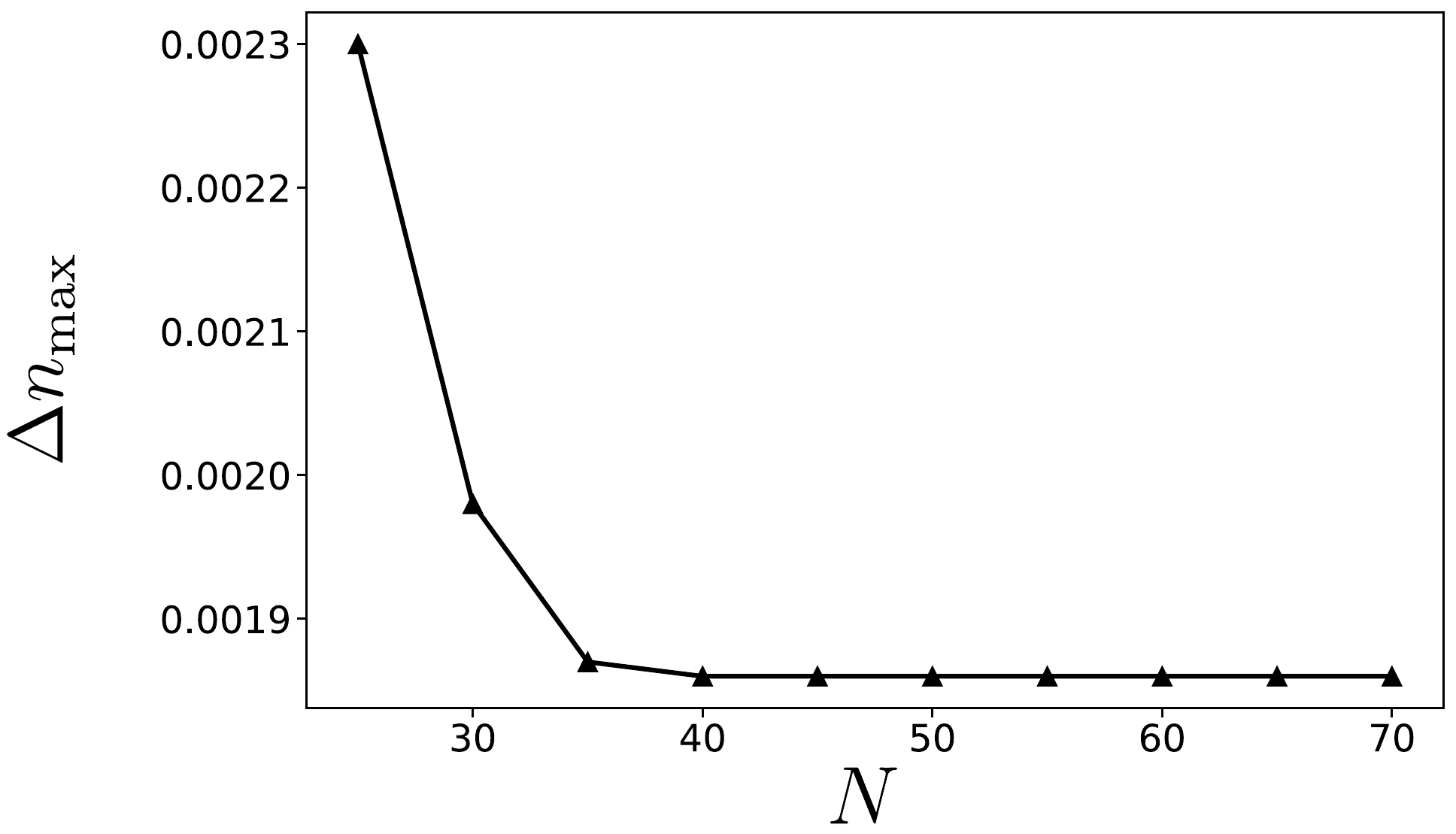}
    \caption{Scaling of the maximum asymmetry $\Delta n_{\mathrm{max}}=n(\mu-2\sigma)-n(\mu+2\sigma)$, for $\mu$ the mean of the initial Gaussian correlation matrix perturbation and $\sigma$ its standard deviation. Here $\mu = \lfloor N/2 \rfloor$ with $N=100$. Also, $\epsilon=7$ in Eq.~\eqref{eq:perturbed_correlation_matrix} and $\sigma=5$, as in Fig.~2 of the main text. We obtain $\Delta n_\mathrm{max}$ by time-evolving the system at each $N$ and finding the time $\tau$ for which the asymmetry is maximal. We clearly see a convergence of $\Delta n_{\mathrm{max}}$ with increasing system size $N$, precluding its interpretation as a finite-size effect.}
    \label{fig:scaling_of_peak_vs_system_size}
\end{figure}
\begin{figure}
    \centering
    \includegraphics[width=0.8\linewidth]{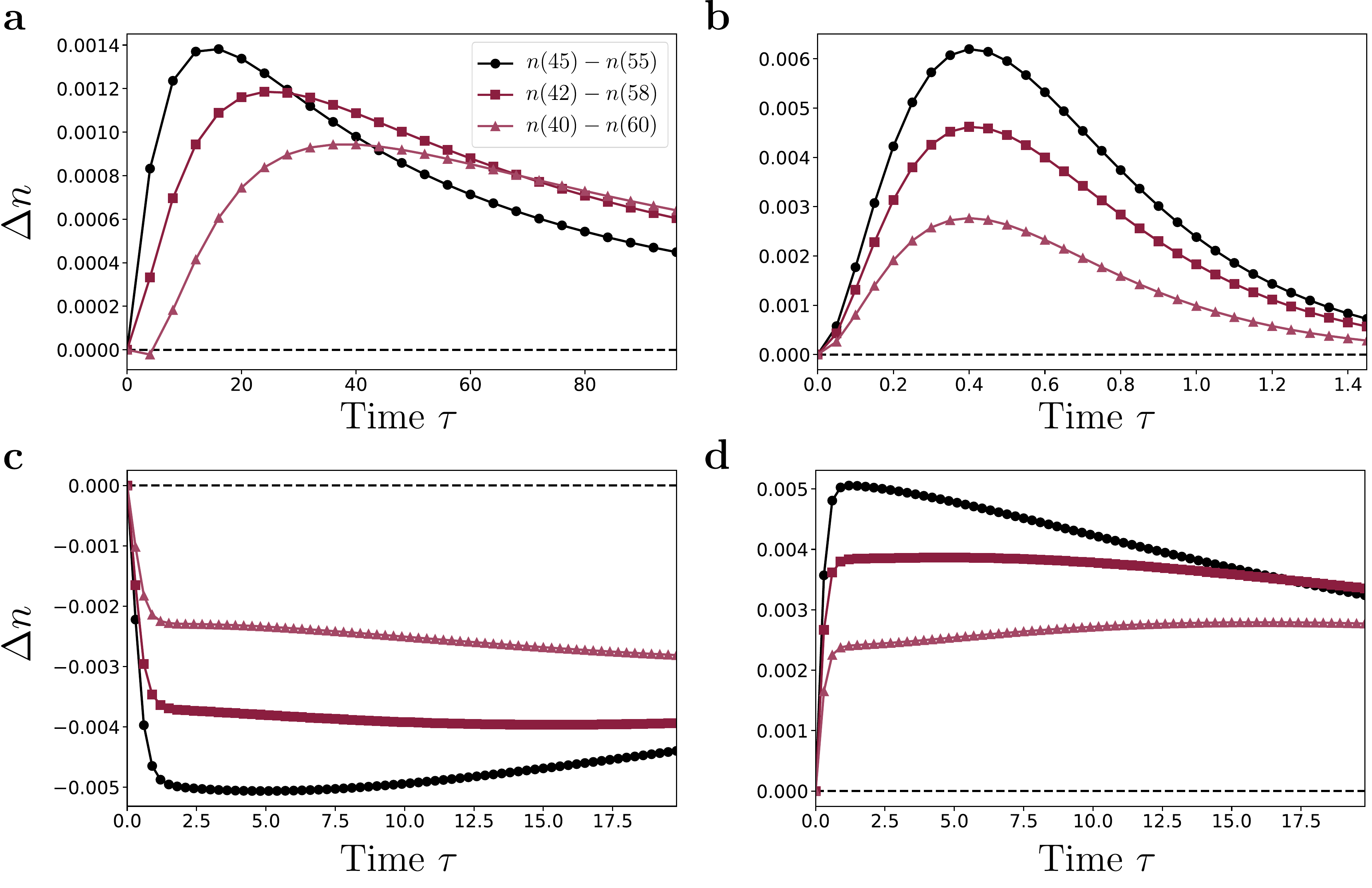}
    \caption{\textbf{a.} Asymmetric diffusion for the case where the decohered point gap winds in the same way as $\Theta$, as in Fig.~\ref{fig:reversed_skin_effect} (with the same parameters).  Correspondingly, the asymmetric diffusion reverses direction compared to Fig.~2 of the main text. \textbf{b.} Postselected dynamics of the same initial perturbation for $N=50$. The timescale is now much shorter, matching the larger magnitude of the decay rate of continuum eigenmodes. We observe non-reciprocal coherent hopping, as in a non-Hermitian Hamiltonian. \textbf{c.} Full dynamics (non-postselected) for a pure-state perturbation atop the steady-state [Eq.~\eqref{eq:pure_state_perturb}] for the same parameters as in Fig.~2 of the main text. The long-lived asymmetric diffusion is the same, with short-lived initial peak explained by coherent hopping, as in panel c. \textbf{d.} Same as panel c but with the reverse-winding parameters as in Fig.~\ref{fig:reversed_skin_effect}.}
    \label{fig:dynamical_sanity_checks}
\end{figure}
We now discuss the behavior of perturbations from the steady state. More precisely, we consider initial states described by correlation matrices
\begin{equation}\label{eq:perturbed_correlation_matrix}
    C(0) = C_S + \epsilon \cdot\delta C,
\end{equation}
for $\epsilon>0$ some small real parameter, and $\delta C$ some perturbation of choice. We will consider $\delta C$ that gives a Gaussian density profile located at the middle of the chain, with some spread $\sigma$.

As established in the main text, one dynamical consequence of the decohered point gap is asymmetric diffusion on a large timescale. A few sanity checks can be performed: firstly, this effect vanishes when $B=0$. Secondly, diffusion is completely symmetric at the points $t=\pm g$, where the decohered point gap closes. Thirdly, the direction of asymmetry reverses with the winding direction of the point gap, as shown in Fig.~\ref{fig:dynamical_sanity_checks}(a). Fourthly, using only the postselected part $\hat{\mathcal{C}}^{\mathrm{post}}$ to time-evolve, this effect disappears and is replaced by asymmetric coherent hopping on a much shorter timescale, as shown in Fig.~\ref{fig:dynamical_sanity_checks}(b).

We verified this for two different initial perturbations $\delta C$. The first is a diagonal matrix with Gaussian density distribution
\begin{equation}
    \delta C_{ii} = e^{-(i-\mu)^2/(2\sigma^2)},
\end{equation}
centered mid-chain, i.e. $\mu = \lfloor N/2 \rfloor$. We also divide this by the maximum argument to normalize. The second is a pure Gaussian wavepacket, constructed from a pure state
\begin{equation}\label{eq:pure_state_perturb}
    \ket{\psi} = \sum_i e^{-(i-\mu)^2/(4\sigma^2)}\ket{i},
\end{equation}
where $i=1,\ldots N$. After normalizing this, we construct $\delta C=\ket{\psi}\bra{\psi}$. This has the same density profile as the other initialization. However, this second initialization kind is not a diagonal correlation matrix and has higher purity. The only difference in their dynamics occurs initially, in the way they couple to continuum eigenmodes. After a few multiples of the continuum timescale, these both display the same qualitative asymmetric diffusion profile, which is near-classical. 

We have also verified that dynamics under $\Chat$ match that obtained from explicit diagonalization of the many-body $4^N\times 4^N$ Lindbladian superoperator for $N=5$ fermions.

\bibliography{refs}